\documentclass[12pt]{article}
\usepackage{jheppub}
\pdfoutput = 1

\usepackage{amsmath}
\usepackage{amssymb}
\usepackage{bbm}
\usepackage{bbold}
\usepackage{braket}
\usepackage{caption}
\usepackage{color}
    \definecolor{darkgreen}{rgb}{0,0.5,0}
    \definecolor{darkred}{rgb}{0.5,0,0}
    \definecolor{darkblue}{rgb}{0,0,0.6}
    \definecolor{purple}{rgb}{0.4,.2,0.7}

\usepackage[mathcal]{eucal}
\usepackage{float}
\usepackage{graphicx}
\usepackage{hyperref}

\usepackage{mathabx}
\usepackage{mathtools}
\usepackage{pdfsync}
\usepackage{slashed}
\usepackage[normalem]{ulem}
\usepackage{upgreek}
\usepackage{url}

\renewcommand\L{Lichnerowicz }

\usepackage[ragged]{footmisc}
\def\be{\begin{equation}}
\def\ee{\end{equation}}

\renewcommand{\tilde}{\widetilde}

\numberwithin{equation}{section}

\begin{document}

\title{Stability of the microcanonical ensemble in Euclidean Quantum Gravity}
\author[a]{Donald~Marolf,}
\author[b]{Jorge~E.~Santos}

\affiliation[a]{Department of Physics, University of California at Santa Barbara, Santa Barbara, CA 93106, U.S.A.}
\affiliation[b]{Department of Applied Mathematics and Theoretical Physics, University of Cambridge, Wilberforce Road, Cambridge, CB3 0WA, UK}

\emailAdd{marolf@ucsb.edu}
\emailAdd{jss55@cam.ac.uk}

\abstract{
This work resolves a longstanding tension between the physically-expected stability of the microcanonical ensemble for gravitating systems and the fact that the known negative mode of the asymptotically flat Schwarzschild black hole decays too rapidly at infinity to affect the ADM energy boundary term at infinity.     The key to our study is that we fix an appropriate {\it off-shell} notion of energy, which we obtain by constructing the microcanonical partition function as an integral transform of the canonical partition function.  After applying the rule-of-thumb for Wick rotations from  our recent companion paper to deal with the conformal mode problem of Euclidean gravity, we find a positive definite action for linear perturbations about any Euclidean Schwarzchild (-AdS) black hole.  Most of our work is done in a cavity with reflecting boundary conditions, but the cavity wall can be removed by taking an appropriate limit.
}

\maketitle

\section{Introduction}
The study of black hole thermodynamics via Euclidean path integrals dates from the seminal work
\cite{Gibbons:1976ue}
of Gibbons and Hawking in the 1970's.  The associated technology for studying the canonical ensemble and analyzing stability is by now well-developed \cite{Gibbons:1976ue,Gibbons:1978ji,Allen:1984bp,Prestidge:1999uq,Dasgupta:2001ue,Kol:2006ga,Headrick:2006ti,Monteiro:2008wr,Monteiro:2009tc,Monteiro:2009ke} and has been applied to a vast range of systems and partition functions (see e.g. \cite{Anninos:2012ft,Benjamin:2020mfz,Cotler:2019nbi,Marolf:2018ldl,Cotler:2021cqa} for recent examples).

In contrast, Euclidean path integral treatments of the microcanonical ensemble have received relatively little attention.  References discussing ``microcanonical'' path integrals and their saddle-points include
\cite{Brown:1992bq,Brown:1993ke,Saad:2018bqo,Marolf:2018ldl}, but these references differ among themselves in their choices of boundary condition and/or partition function to compute.  Furthermore, the only reference we have found addressing stability of saddle points for microcanonical path integrals in Einstein-Hilbert gravity is \cite{Allen:1984bp}, which appears to claim that Euclidean Schwarzschild with zero cosmological constant is an unstable saddle in a reflecting cavity of any size.
Since Schwarzschild is known to be a local maximum of bifurcation-surface area under fixed-energy on-shell perturbations,\footnote{For zero cosmological constant this local maximum follows in particular from the classic analyses \cite{Regge:1957td,Zerilli:1970se} of dynamical stability of Schwarzschild and the fact \cite{Hollands:2012sf} that such stability is equivalent positivity of the canonical 2nd order energy ${\cal E}= \delta^2 M - \frac{\kappa}{8\pi G} \delta^2 A -\sum_A \Omega_A \delta^2 J_A$ with respect to perturbations that preserve the energy at first order (since the canonical energy turns out to depend only on the first-order part of the perturbation and since Schwarzschild has $\Omega_A=0$).} such a result would be in strong tension with the interpretation of black holes as thermodynamic systems with entropy $S_{BH} = \frac{A}{4G}$.

In fact, the tension goes rather deeper.  While the Euclidean 
asymptotically flat Schwarzschild black hole is a saddle for the
the canonical-ensemble path integral, in that context it has a negative mode that  vanishes rapidly at infinity.  As a result, the negative mode does not affect the boundary term that computes the asymptotically-flat ADM energy, and fixing this boundary term cannot remove the negative mode.  For fixed black hole mass $M$, the negative mode should thus again persist in a large enough cavity or in an asymptotically AdS space of sufficiently large AdS scale $\ell$. Though we have not found explicit discussion of this point in published works, we believe it to be the reason for the dearth of literature on microcanonical stability in gravitational systems.

We resolve this apparent contradiction below by following the approach to microcanonical gravitational path integrals described in \cite{Marolf:2018ldl} (see also related comments in \cite{Saad:2018bqo}).  This approach builds on the success of the standard path integral for the canonical partition function
\begin{equation}
\label{eq:can}
Z(\beta) = \int {\cal D}{\sf g}_{R,\beta} e^{-I[g]}
\end{equation}
in a cavity of radius $R$.  In particular, it does so by {\it defining} the microcanonical partition function $Z_{micro}(E)$ as an integral of $Z(\beta)$ over an appropriate contour $\Gamma$ in the complex $\beta$-plane; \emph{i.e.}

\begin{equation}
\label{eq:micro}
Z_{micro}(E) = \int_{\Gamma} d \beta \ F_E(\beta) \ Z(\beta) = \int_{\Gamma} d \beta \int {\cal D}{\sf g}_{R,\beta} \ F_E(\beta) e^{-I[g]}.
\end{equation}
Here $F_E(\beta)$  is a weighting function and ${\cal D}{\sf g}_{R,\beta}$ denotes an appropriate measure on the space of metrics with boundary $S^n \times S^1$ in which the $S^n$ has radius $R$ and the $S^1$ has proper length\footnote{Since we will generally work in a finite cavity, it is natural to parameterize the size of the $S^1$ using proper distance and to parameterize the size of the $S^n$ using the area-radius.  It is straightforward to rescale by an appropriate function of $R$ in order to, for example, instead use a notion of distance defined by a boundary metric in some conformal frame when taking the $R\rightarrow \infty$ limit in asymptotically AdS spacetimes.} $\beta$.  As described in \cite{Marolf:2018ldl}, with an appropriate $F_E(\beta)$ classical solutions with energy $E$ will provide saddle points for the path integral \eqref{eq:micro}.  The details of $\Gamma$ and $F_E(\beta)$ will be specified below.

The starting point for the current work is the observation that one can  perform the integral over $\beta$ in \eqref{eq:micro} {\it before} evaluating the integral over $g_{\beta, R}$. The result is a microcanonical path integral that fixes a certain {\it off-shell} notion of gravitational energy which includes non-trivial contributions from degrees of freedom that violate the gravitational constraints.  As a result, our energy is not just a boundary term at the cavity wall.   The details of this notion of energy depend on which aspects of the metric are held fixed when integrating over $\beta$, but for a simple choice and for static fluctuations it becomes essentially the standard off-shell Hamiltonian associated with translations along the bulk Killing field.  In addition, for {\it linear} static fluctuations about AdS-Schwarzschild it becomes equivalent to $1/4G$ times the area $A$ of the extremal surface at the Euclidean horizon.   

We analyze this choice in detail and, using the rule-of-thumb from \cite{CanoPaper} for Wick rotations to deal with the conformal mode problem of the Euclidean path integral, we find the microcanonical Euclidean action at quadratic order to be positive definite.  Thus, as expected, all Euclidean (AdS) Schwarzschild black holes provides stable saddles for our microcanonical path integral.  This provides an important check on the admittedly-formal reasoning we use to derive our microcanonical path integral.

We begin by describing the general construction of our microcanonical path integral in section \ref{sec:gen}.  This involves specifying the details of the various ingredients that compose \eqref{eq:micro}, performing the integral over $\beta$, and determining the appropriate off-shell notion of energy that becomes fixed. As stated above, with simple choices and for static linearized fluctuations about AdS-Schwazschild saddles this off-shell energy becomes the area of the extremal surface at the Euclidean horizon, though we also discuss 2nd order contributions. 

We then describe the strategy to be used to investigate saddles of the resulting path integral in section \ref{sec:SL}.
The actual analysis of stability then begins in section \ref{sec:perts}, which considers Euclidean AdS$_d$-Schwarzschild black holes in a cavity for $d=4,5$ and describes details of their perturbations. Stability results are presented in section \ref{sec:res}. As indicated above, the numerical results show these solutions to define stable saddles of our microcanonical path integral.  The case of vanishing cosmological constant is included as a natural limit.  We conclude with a summary and discussion of open questions in section \ref{sec:disc}.

\section{Microcanonical path integrals and their fluctuations\label{sec:gen}}

Let us consider the canonical path integral \eqref{eq:can} computing the partition function $Z(\beta)$ for a gravitating system with Euclidean boundary $S^n \times S^1$.  We take the $S^1$ (which we think of as Euclidean time) to have proper length $\beta$ and we take the sphere $S^n$ to have area-radius $R$.  Thus we choose to work in a finite cavity.  This choice plays little role in the present section, and indeed it is trivial to take $R \rightarrow \infty$.
While it is far from clear that the finite $R$ setting can define a consistent theory at the microscopic level (see e.g. \cite{Witten:2018lgb,Andrade:2015qea,Andrade:2015gja}), we make this choice here for consistency with later sections where the finite cavity provides a useful infrared (IR) regulator for numerical calculations.  In particular, it renders the spectrum of appropriate differential operators discrete.

As mentioned in the introduction, we will follow the approach to the microcanonical path integral described in \cite{Marolf:2018ldl}.  To explain the idea, recall that
one expects the canonical path integral \eqref{eq:can} to in some sense compute a statistical sum $z(\beta): = \sum_E e^{-\beta E}$, where the sum is over all energy levels $E$ of the gravitational system.  Analytically continuing $\beta \rightarrow \beta +iT$ with $\beta, T$ real, choosing an arbitrary reference value $\beta_0$,  and introducing the weighting function $f_{E_0}(T)$, we may consider the integral
\begin{align}
z_{micro}(E_0) :&= e^{\beta_0 E_0} \int \mathrm{d}T\,z(\beta_0 +iT) f_{E_0}(T) \nonumber
\\
& = \sum_E  e^{-\beta_0(E_0-E)} \tilde f_{E_0}(E).
\label{eq:statmicro}
\end{align}
Note that this is of much the same form as \eqref{eq:micro} with $f_{E_0}(T) = F_{E_0}(\beta_0 + iT)$ and $\Gamma$ given for all $E$ by the contour ${\rm Re}\   \beta = \beta_0$. We see that the result is a microcanonial statistical model associated with the Fourier-transformed weighting function $\tilde f_{E_0}(E)$ (where we have made an appropriate choice of normalization for the Fourier transform). It is then natural to choose $\tilde f_{E_0}(E)$ to be real and to be concentrated in a small window $\Delta E$ near $E_0$, in which case we find 
\begin{equation}
\label{eq:statmicro2}
z_{micro}(E_0) \approx \sum_{E \in \Delta E}  (1) = {\rm{number \ of \ states \ with}} \ E\in\Delta E.
\end{equation}

We may thus make the analogous definition for our gravitational path integral, taking the contour $\Gamma$ to be independent of $E$ and to run parallel to the imaginary $\beta$ axis with ${\rm Re}   \ \beta = \beta_0$ fixed.  In particular, we  define
\begin{equation}
\label{eq:micro2}
Z_{micro}(E_0) = e^{\beta_0 E} \int_{T\in {\mathbb R}} \mathrm{d} T \int {\cal D}{\sf g}_{R,\beta_0+iT}\,f_{E_0}(T) e^{-I[g]}.
\end{equation}
Ref. \cite{Marolf:2018ldl} took $\tilde f_{E_0}$ to be a Gaussian centered at $E_0$ and studied the limit where the width $\sigma$ of this Gaussian was small.  We will do the same here, choosing
\begin{equation}
\label{eq:fchoice}
f_{E_0}(T) = e^{-iE_0 T} e^{-\sigma^2 T^2/2G} =  e^{-ie_0 T/G} e^{-\sigma^2 T^2/2G},
\end{equation}
where we take $E_0= e_0/G$ with $e_0$ and $\sigma^2$ independent of the Newton constant $G$ so that all dependence on $G$ is explicit on the right-hand side of \eqref{eq:fchoice}.
In the limit of strictly-vanishing $\sigma$ this construction is equivalent to performing an inverse Laplace transform of the canonical partition function $Z(\beta)$.  Saddle points for the resulting path integral were discussed in \cite{Marolf:2018ldl}.

We wish to extend the discussion of \cite{Marolf:2018ldl} to consider off-shell fluctuations about such saddles.  One approach would be to perform all integrals simultaneously using the saddle-point approximation and to study fluctuations that allow changes in both $T$ and ${\sf g}_{R,\beta_0+iT}$.  Recall, however, that has been useful in the past to formulate the stability problem in terms of  a Sturm-Liouville-type eigenvalue problem.  In particular, this is the setting required to apply the rule-of-thumb from \cite{CanoPaper} to specify a contour of integration for the path integral. However, this will not be the case if we study all fluctuations together as the $T$ variable adds a degree of freedom, effectively removing one of the two boundary conditions that one would expect.  We will thus perform the integral over $T$ {\it before} integrating over metrics with fixed $T$.  This allows us to write the result as a path integral over metrics which, in the limit $\sigma \rightarrow 0$ and for time-independent metrics, can be described as having fixed value of a off-shell energy that we call $H$. The stability problem can then be formulated in a more familiar way.

In order to implement the above procedure, we must first introduce an additional structure on the space of metrics that defines what is meant by `metrics with fixed $T$'. This is done in section \ref{sec:org} below.  We also henceforth restrict consideration to time-independent metrics.  In particular, though we will continue to use the same notation ${\cal D} {\sf g}_{R, \beta}$ for the functional integral, this symbol should henceforth be understood to be a measure on the space of time-independent metrics.  We hope to provide a more complete treatment in future work by using the Hamiltonian formalism.  Section \ref{sec:sandf} will then describe properties of the saddles and fluctations about saddles in the microcanonical ensemble that will finally allow us to formulate microcanonical stability as a Sturm-Liouville-type problem in section \ref{sec:SL}.
\subsection{The microcanonical the gravitational Path Integral}
\label{sec:org}
As noted above, our approach requires us to introduce a new structure on the space of (time-independent) metrics in order to define what is meant by `metrics with fixed $T$'.
Specifically, let us consider the space ${\mathfrak G}_R$ of (time-independent) metrics which induce boundary metrics $S^1 \times S^{d-2}$ on the cavity walls.  We take this to be a metric product so that the $S^1$ is orthogonal to the $S^{d-2}$ and the latter is a  round sphere.  The subscript $R$ indicates the radius of the $S^{d-2}$ but the proper length of the $S^1$ factor is not constrained.

\begin{figure}[h]
\centering\includegraphics[width=.6\linewidth]{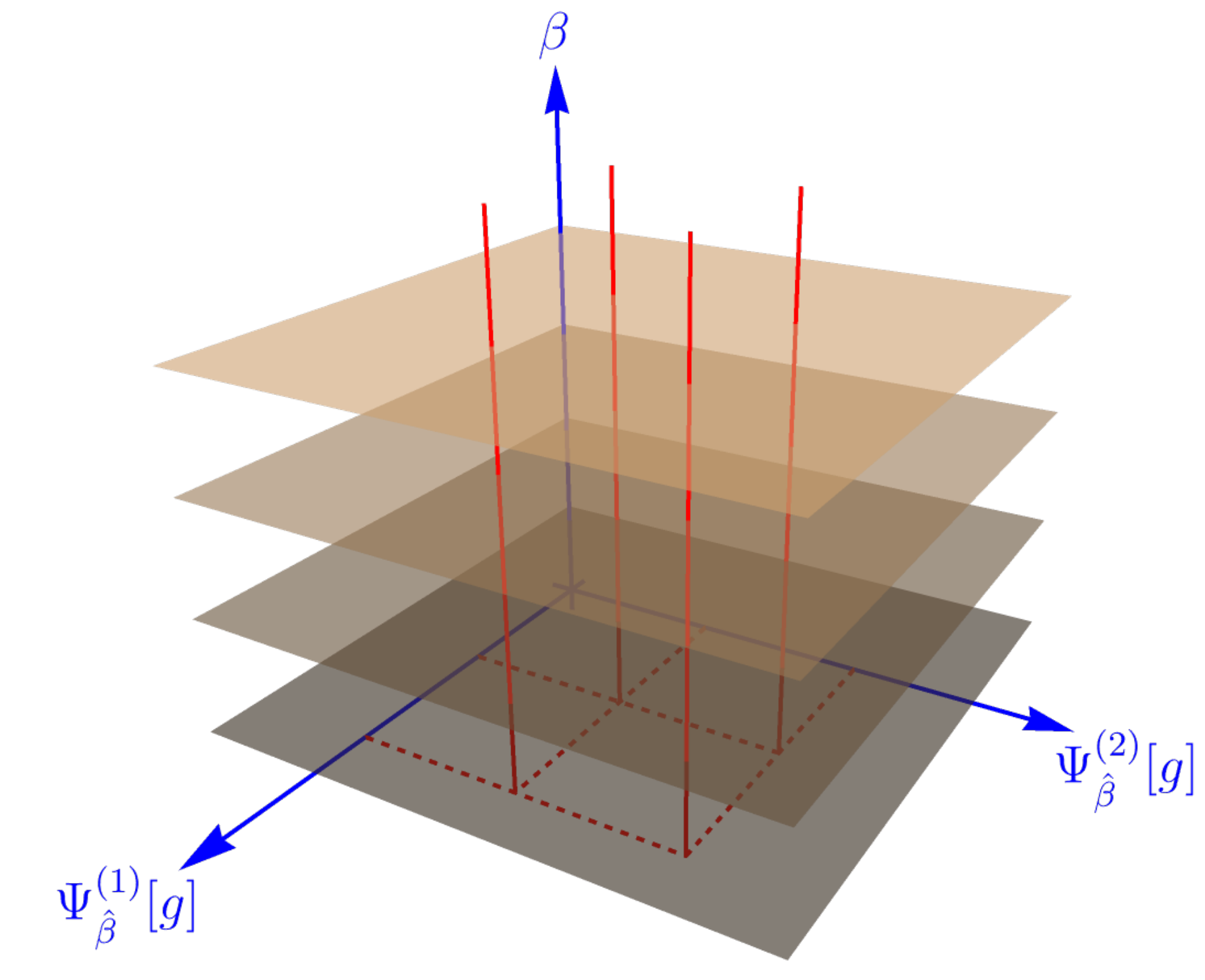}
\caption{An illustration of how we organize the space of bulk metrics ${\mathfrak G}_R$.  The horizontal surfaces are codimension-1 slices of constant $\beta$.  The red lines are 1-dimensional surfaces that define a notion of what it means to keep `the rest of the metric' constant while changing $\beta$, and can thus be said to define additional coordinates  on the space of all metrics.  We may in particular proceed by choosing a reference value $\hat \beta$ and lifting any coordinates $\Psi^{(1)}$, $\Psi^{(2)}$ on the corresponding horizontal surface to the full space by taking them to be constant along the red lines.  We use $\Psi_{\hat \beta}[g]$ to denote the result of inverting this procedure by taking a general metric $g$ and `projecting' it along the red lines into the horizontal surface at  $\hat \beta$.}
\label{fig:slicing}
\end{figure}

One expects the physics to be independent of how we divide the integral over metrics into an integral over $T$ and integrals over `metrics with constant $T$'.   At least on the surface, changing this choice just corresponds to performing the required integrals in a different order.   However, we note that the microcanonical path integral \eqref{eq:micro2} integrates over a contour on which metrics cannot be real, and that specifying precisely what is meant by allowing the period $\beta = \beta_0+iT$ to be complex `without deforming the contour for the other variables of integration variables' requires us to make some such choice.    Thus the expectation that predictions of the ensemble will not depend on this choice relies on assuming that the contour can be deformed without changing the partition function. While this seems entirely plausible, we will not investigate it further in this work.  Instead, we simply make a particular choice below and leave further exploration of this issue for the future.

Our choice will be specified by using a set of coordinates $\tau,x^i$ with $\tau$ singled out as describing (Euclidean) time. For all spacetimes we take $\tau$ to be periodic with period $2\pi$. For simplicity, we will in fact choose coordinates so that the induced metric on the cavity wall takes the form
\begin{equation}
\mathrm{d}s^2_{\partial} = - \frac{\beta^2}{4\pi^2} \mathrm{d}\tau^2 + R^2 \mathrm{d}\Omega^2_{d-2}\,,
\label{eq:induced}
\end{equation}
where $\mathrm{d}\Omega_{d-2}^2$ is the line element on a round $S^{d-2}$ sphere with unit radius. We also define $x^i = (r, \Omega)$ for an appropriate set of angles $\Omega$ on $S^{d-2}$, and such that the cavity wall lies at $r=R$. Moreover, the $\beta$ appearing in Eq.~(\ref{eq:induced}), which we will call the period of the $S^1$, is independent of $x^i,\tau$. Furthermore, there will be some codimension-2 surface $\gamma$ at which the $\tau$ circle shrinks to zero size.  It is natural to think of $\gamma$ as being a Euclidean black hole horizon.

Given a (time-independent) metric $\hat g\in {\mathfrak G}_R$ with length $\hat \beta$ for the $S^1$ factor at the cavity wall, we will define the metric $\Psi_{\beta}[\hat g]$ via the scaling:
\begin{eqnarray}
\label{eq:relate}
(\Psi_{\beta}[\hat g])_{\tau \tau}(x^i)  &=&  \left(\frac{\beta}{\hat \beta}\right)^2 \hat g_{\tau \tau}(x^i), \ \ \ \cr
(\Psi_{\beta}[\hat g])_{\tau i}(x^i)    &=&  \frac{\beta}{\hat \beta}\hat g_{\tau i}(x^i), \ \ \ \cr
(\Psi_{\beta}[\hat g])_{ij}(x^i) &=&  \hat g_{ij}(x^i).
\end{eqnarray}
Note that since the period of $\tau$ remains $2\pi$ instead of becoming $2 \pi \hat \beta/\beta$, \eqref{eq:relate} does {\it not} describe a spacetime diffeomorphism associated with rescaling $\tau$. As a result, the map $\Psi_\beta$ can change the local spacetime geometry at the surface $\gamma$ where the $\tau$ circle shrinks to zero size.  Indeed, when $g$ is smooth the metric $\Psi_{\beta}[\hat g]$ will have a conical singularity at $\gamma$ for all $\beta \neq \hat \beta$.  We will thus explicitly include metrics with such conical singularities\footnote{\label{foot:def} We tentatively define conical singularities at $\gamma$ as those singularities generated by applying $\Psi_{\beta}$ to smooth metrics $\hat g$ for all values of $\hat \beta$.  It is an interesting question whether this is really the most appropriate notion of such a definition, or whether one might wish to instead impose a definition along the lines of the appendices in \cite{Dong:2019piw}.  The current work is sufficiently formal that we will not concern ourselves with this issue further, though it would be good to address this in the future.} in the space of metrics ${\mathfrak G}_R$, though  as usual, we will not attempt to specify in detail the space of metrics over which the path integral will be performed.  We simply imagine that ${\mathfrak G}_R$ can be constructed as an appropriate completion of the space of smooth metrics, so that properties of the singular members of ${\mathfrak G}_R$ can be derived by taking limits of results for the smooth case\footnote{In particular, while it is traditional to search for negative modes around saddles for the canonical ensemble by studying smooth fluctuations, treating conical singularities as limits of smooth metrics means that allowing this conical singularity to fluctuate can add nothing new.}.  This is natural for the above conical geometries, and also for geometries that may involve conical singularities on other codimension-2 surfaces.

Nevertheless, the idea that metrics in the domain of integration are smooth is a convenient and familiar fiction.  In order to preserve this fiction to the greatest extent possible, we now define another projection $\tilde \Psi$ on ${\mathfrak G}_R$ which removes conical singularities at $\gamma$.  In particular, if $\hat g \in {\mathfrak G}_R$ has a conical singularity on $\gamma$, we define $\tilde \Psi[\hat g] = \Psi_{\tilde \beta}[\hat g]$ for the particular value of $\tilde \beta$ that removes the conical singularity at $\gamma$.   As described in footnote \ref{foot:def}, we have defined conical singularities to be those for which there is one (and only one) value of $\tilde \beta$ that achieves this goal.  We will assume that we can neglect any other singularities of metrics $g\in {\mathfrak G}_R$ in the same way that classic treatments
 of stability in the canonical ensemble \cite{Gibbons:1976ue,Gibbons:1978ji,Prestidge:1999uq} focus on smooth off-shell perturbations and describe more singular perturbations by taking limits.

 Below, we use $\tilde {\mathfrak G}_R$ to denote the image $\tilde \Psi[{\mathfrak G}_R]$, which we assume is just the subspace of ${\mathfrak G}_R$ where there is no conical singularity on $\gamma$.  We similarly use $\tilde {\sf g}_R$ to denote an appropriate set of coordinates on $\tilde {\mathfrak G}_R$.
 Our microcanonical partition function \eqref{eq:micro2} can thus be written
\begin{equation}
\label{eq:micro3}
Z_{micro}(E) = e^{\beta_0 E} \int_{T\in {\mathbb R}} d T \int_{\tilde g \in \tilde {\cal G}} {\cal D}\tilde {\sf g}_{R} \ \mu({\sf g}_R, T) \  f_E(T) e^{-I[\Psi_{\beta_0 + iT}(\tilde g)]},
\end{equation}
where $\mu({\sf g}_R, T)$ is the Jacobian associated with the change of coordinates on ${\mathfrak G}_R$ from $(\beta, {\sf g}_{R, \beta})$ to $(\beta, \tilde {\sf g}_R)$.   So long as this Jacobian is non-singular it can be ignored in the semiclassical limit studied here.  We will thus simply set $\mu=1$ in all formulae below\footnote{\label{foot:measure}This is admittedly somewhat dangerous as deriving the covariant path integral from the Hamiltonian formalism suggests that the dependence of the measure on $\beta$ will in fact be highly singular at $\beta=0$.  A better approach is to follow a Hamiltonian analysis,  which in particular gives full control over the measure.  This will be explored in a subsequent paper. Here we will simply use the above very formal analysis to motivate our proposal for the microcanonical path integral, and for a stability criterion for its saddles, which we then test in section \ref{sec:res}.}.

We are now ready to perform the integral over $T$ in \eqref{eq:micro2}. 
The key point is that our restriction to time-independent metrics implies the action $I[\Psi_{\beta}(\tilde g)]$ to be linear in $\beta$.  Indeed, in $\tau, x^i$ coordinates the action can be written as a sum of a $\beta$-independent boundary term at $\gamma$ and an integral over $\tau, x^i$ (where we take the latter to include the appropriate boundary terms at the cavity wall).  Both terms are invariant under diffeomorphisms of $\tau, x^i$.  But the transformation \eqref{eq:relate} differs from a diffeomorphism by failing to rescale $\tau$ itself, and in particular by leaving the period of $\tau$ fixed.  As a result, the integral over $\tau$ is in fact proportional to $\beta = \beta_0 + iT$ at fixed $\tilde g$.  We thus write

\begin{equation}
\label{eq:linact}
I[\Psi_{\beta}(\tilde g)] = -\frac{A(\tilde g)}{4G}+ \beta H[\tilde g].
\end{equation}
For Einstein-Hilbert gravity with minimally-coupled matter, the appropriate $A(\tilde g)$ at $\gamma$ is well-known to be just the area of $\gamma$; see e.g \cite{Banados:1993qp}.  In more general theories of gravity it will be a corresponding geometric entropy term, but for simplicity we continue to denote it by $A(\tilde g)$.

While we have simply defined $H[\tilde g]$ as the coefficient of $\beta$ in \eqref{eq:linact}, we will refer to this quantity as the off-shell Euclidean Hamiltonian (or as just the Hamiltonian when there is no possibility of confusion).   This terminology is natural since, if we happen to evaluate $\partial_\beta I|_{\tilde g}$ on-shell, then stationarity of $I$ guarantees that all bulk contributions will vanish.  Using either a boundary stress tensor or a Hamilton-Jacobi definition of energy, the result then reduces to the usual on-shell Hamiltonian (as defined by the whatever boundary terms we may have chosen to use in defining $I$).  Due to our restriction to time-independent metrics, this $H$ is time-independent even off-shell.

Furthermore, by comparing with the Hamiltonian form of the gravitational action  (see e.g. \cite{Banados:1993qp}) and noting that the usual symplectic term $\pi^{ij} \dot{h}_{ij}$ vanishes due to time-independence of the metric, we see that the bulk terms in $H[\tilde g]$ are precisely the usual (Euclidean) Hamiltonian and momentum constraints for $g = \Psi_\beta \tilde g$  multiplied by corresponding (Euclidean) lapse and shift and integrated over the spacetime.  Note that the Euclidean lapse is complex when our $T$ is real.   Corresponding statements will also be true for more complicated theories with higher derivative corrections.

In the approximation where we neglect dependence of $\mu$ on $T$, it is now straightforward to perform the $T$-integral in \eqref{eq:micro3}.  The integral simply enacts a fourier transform:
\begin{equation}
\label{eq:micro4}
Z_{micro}(E) \approx e^{\beta_0 E} \int_{\tilde g \in \tilde {\cal G}} {\cal D}\tilde {\sf g}_{R} \   \tilde f_E(H[\tilde g]) e^{A(\tilde g)/4G},
\end{equation}
In the limit $\sigma^2 \rightarrow 0$, up to normalization the effect of the factor $\tilde f_E(H)$ is simply to restrict the domain of integration metrics $\tilde g \in {\mathfrak G}_R$ which also satisfy $H[\tilde g] = E$.
This is the microcanonical path integral whose stability we will study in section \ref{sec:res}, where we will find agreement with physical expectations.

In summary, with our restriction to time-independent metrics and modulo footnote \ref{foot:measure}, performing the integral over $T$ and taking $\sigma^2 \rightarrow 0$ gives a microcanonical path integral  that integrates over metrics $\tilde g \in \tilde {\mathfrak G}_R$  with $H[\tilde g] = E$.  Much as one might naively expect, this is an integral over smooth metrics (or at least those without conical singularities at $\gamma$ since $\tilde g \in \tilde {\mathfrak G}_R$) with fixed (off-shell) energy, but where the period of the $S^1$ is allowed to fluctuate.
Furthermore, since the integral over $T$ was {\it precisely} the Fourier transform of a Gaussian, fluctuations in $T$ about the saddle-point are clearly stable.

We now wish to evaluate the remaining integrals using the saddle-point approximation.  For Einstein-Hilbert gravity the Euclidean (AdS)-Schwarzschild solutions all give saddles, and we will see in section \ref{sec:res} that with cosmological constant $\Lambda \le 0$ these saddles are all stable at quadratic order. However, before embarking on explicit computations, we first provide a few further results in section \ref{sec:sandf} that will allow us to recast the analysis of perturbative stability as a standard search for negative modes in a Sturm-Liouville-type problem in section \ref{sec:SL}. 

\subsection{The microcanonical action: Saddles and fluctuations}
\label{sec:sandf}

From \eqref{eq:micro4}, we can identify $(-A/4G)$ as the microcanonical action for time-independent metrics.  We wish to study saddle points that make $(-A[\tilde g]/4G)$ stationary under variations $\delta \tilde g$ that preserve the constraint $H[\tilde g]=E$, or in other words for which $\delta H=0$.  This makes it easy to relate such saddles to stationary points $\tilde g_*$ of the canonical ensemble variational problem defined by varying $I[\tilde g] = -A/4G + \tilde \beta H[\tilde g]$ at fixed $\tilde \beta$.  Indeed, under a general variation of $\delta \tilde g$ within $\tilde {\mathfrak G}_R$ about a time-independent background $\tilde g_*$, for any constant $E$ we have the variation

\begin{equation}
\label{eq:relatesaddles}
\delta (I-\tilde \beta E) = - \delta A/4G + (H[\tilde g_*]-E) \delta \tilde \beta  + \tilde \beta \delta H.
\end{equation}
Note that the constant $E$ has so far been taken to be arbitrary and need not be related in any way to $\tilde g_*$ or $\delta \tilde g$.

To see the implications of \eqref{eq:relatesaddles}, note that if $\tilde g_*$ is a saddle for the canonical ensemble variational problem with period $\tilde \beta_*$, then the left-hand-side must be proportional to $\delta \beta$.  But the area $A$ and energy $H$ are both manifestly independent of $\beta$ when $\Psi_{\hat\beta}[\tilde g]$ is held fixed for any $\hat \beta$.  Thus from the right-hand-side we see that the result is proportional to $H[\tilde g_*]-E$, and in particular that it vanishes if we choose $E = H[\tilde g_*]$. The remaining terms on the right-hand-side must therefore cancel among themselves, giving us an off-shell version of the first law of black hole mechanics\footnote{The derivation of \eqref{eq:1st} made critical use of the fact that $g_*$ has a Euclidean `time-translation' symmetry whose orbits shrink to zero size at the  Euclidean horizon.  For rotating black holes with angular velocity $\Omega_H$, this would be true for the horizon-generating Killing field $\chi = \partial_t + \Omega_H\, \partial_\phi$.  So, as a result of the extra $(-)$ sign in the conventional definition of energy, in that context the quantity we call $H$ would be better thought of as $H-\Omega_H\, J$.},
\begin{equation}
\label{eq:1st}
\delta A/4G = \tilde \beta_* \delta H,
\end{equation}
where \eqref{eq:1st} is valid for first-order fluctuations $\tilde g$ about a classical solution $\tilde g_*$.
In particular, we see that the microcanonical action $A/4G$ is stationary at $g_*$ under variations that preserve $H$.  Thus the saddles for the canonical and microcanonical problems coincide.

We would now like to understand fixed-$H$ variations at quadratic order.  Since the constraint $\delta H =0$ is non-trivial to impose, we will study it order-by-order in perturbation theory about $g_*$.  The argument above shows that, at first order in $\delta \tilde g$, we may in fact replace it by the simpler constraint $\delta A = 0$.

Let us thus suppose that $\delta \tilde g = \delta \tilde g_1 + \delta \tilde g_2 + \dots$ with $\delta \tilde g_k$ of order $(\delta \tilde g_1)^k$ and with $\delta \tilde g_1$ chosen to preserve $H$ at linear order; \emph{i.e.},
\begin{equation}
\label{eq:H1v}
\delta_1 H : = H[\tilde g + \delta \tilde g_1] - H[\tilde g] = O(\delta \tilde g_1)^2.
\end{equation}
Furthermore, because the area $A$ of the surface $\gamma$ is a relatively simple function of the metric, let us in fact choose $\delta \tilde g_1$ to preserve $A$ exactly:
\begin{equation}
\label{eq:fixA}
A[g_* + \delta \tilde g_1] = A[g_*].
\end{equation}
Note that the choice \eqref{eq:fixA} is consistent with \eqref{eq:H1v} due to \eqref{eq:1st}.

We now expand
\begin{equation}
\label{eq:Hvars}
\delta H = \int \frac{\delta  H}{\delta \tilde g} \delta \tilde g + \frac{1}{2}\int \int' \frac{\delta^2  H}{\delta' \tilde g \delta \tilde g} \delta \tilde g \delta' \tilde g + O(\delta g^3),
\end{equation}
where we have suppressed all indices and we have written $\delta \tilde g, \delta' \tilde g$ for the variations of $\tilde g$ at $x$ and $x'$.  The symbol $\int'$ denotes an integral over $x'$.  Here all functional derivatives are evaluated at the background solution $\tilde g_*$.

In the notation of \eqref{eq:Hvars}, the condition \eqref{eq:fixA} takes the form
\begin{equation}
\label{eq:H1d}
\int \frac{\delta H}{\delta g} \delta g_1 =0.
\end{equation}
To obtain $\delta H =O(\delta \tilde g_1^3)$ we must therefore have
\begin{equation}
\label{eq:H2}
\int \frac{\delta  H}{\delta \tilde g} \delta \tilde  g_2 = -  \frac{1}{2} \int \int' \frac{\delta^2 H}{\delta' \tilde g \delta \tilde g} \delta \tilde g_1 \delta' \tilde g_1 + O(\delta g_1^3).
\end{equation}

We now wish to evaluate
$A(\tilde  g_{*} + \delta \tilde  g) - A(g_{*})$ to second order in $\delta_1 g$.  Using \eqref{eq:fixA} and the fact that $\delta \tilde g_2 = O(\delta \tilde g_1^2)$ we find
\begin{eqnarray}
\label{eq:moreAvars}
 A(\tilde g_{*} + \delta \tilde g_1 + \delta \tilde g_2) - A(g_{*}) &=&  A(\tilde g_{*} + \delta \tilde g_1 + \delta \tilde g_2) - A(g_{*}+ \delta  \tilde g_1)\nonumber
 \\ &=& \int \frac{\delta A}{\delta  \tilde g}|_{( \tilde g_{*}+ \delta  \tilde g_1)}\delta \tilde g_2 +O[(\delta  \tilde g_2)^2] \nonumber
 \\
 &=& \int \frac{\delta A}{\delta  \tilde g}|_{\tilde g_*} \delta \tilde  g_2 +O[(\delta \tilde  g_1)^3].
\end{eqnarray}
In the final step we have used the fact that $\frac{\delta A}{\delta  \tilde g}|_{\tilde g_*}$ differs from $\frac{\delta A}{\delta  \tilde g}|_{( \tilde g_{*}+ \delta  \tilde g_1)}$ by a term of order $\delta g_1$, which when multiplied by by $\delta g_2$ then gives a term of order $(\delta g_1)^3$.

Now,  by \eqref{eq:1st} we may replace the functional derivatives of $A$ in the last line with $4G\tilde \beta_* $ times the functional derivatives of $H$. We thus find

 \begin{eqnarray}
 \label{eq:deltaAasdeltaH}
 -\frac{\delta A}{4G}  &=&  - \tilde \beta_* \int \frac{\delta H}{\delta \tilde  g}|_{\tilde g_*} \delta \tilde g_2 +O[(\delta \tilde g_2)^3],\nonumber
\\
&=&  +  \frac{ \tilde \beta_*}{2} \int \int' \frac{\delta^2 H}{\delta' \tilde g \delta \tilde g} \delta  \tilde  g_1 \delta' \tilde  g_1 + O[(\delta \tilde g_1)^3],
\end{eqnarray}
where the last equality uses \eqref{eq:moreAvars}.

In particular since $\tilde \beta_* >0$, 
for fluctuations that preserve $H$ we see that the microcanonical action $-A/4G$ is postive definite at second order precisely when fluctuations that preserve $A$ give a positive definite quadratic term in $H$.  That is to say, microcanonical saddles are stable if and only if they describe a local minimum of the energy at fixed $A[\gamma]$.  Since we expect $\frac{A[\gamma]}{4G}$ to play the role of entropy, this is a physically sensible conclusion.

In our analysis below, it will be useful to further relate \eqref{eq:deltaAasdeltaH} to the usual Euclidean action $I$ of the canonical ensemble.  This is straightforward, since for fluctuations $\delta_1 \tilde g$ that preserve $A$ we have

\begin{eqnarray}
\int \int' \frac{\delta^2 I}{\delta' \tilde g \delta \tilde g} \delta_1 \tilde g \delta'_1 \tilde g  &=& \tilde \beta_* \int \int' \frac{\delta^2  H}{\delta' \tilde g \delta \tilde g} \delta_1 \tilde g \delta'_1 \tilde g \nonumber
\\&+& \int \int' \frac{\delta H}{\delta' \tilde g}\frac{\delta \beta }{\delta \tilde g} \delta_1 \tilde g \delta'_1 \tilde g  \nonumber
\\ &+& \int \int' \frac{\delta  \tilde \beta }{\delta' \tilde g}\frac{\delta H}{\delta \tilde g} \delta_1 \tilde g \delta'_1 \tilde g \nonumber
\\ &+& H \int \int' \frac{\delta^2 \tilde \beta}{\delta' \tilde g \delta \tilde g} \delta_1 \tilde g \delta'_1 \tilde g .
\end{eqnarray}
Note that the middle two terms vanish when evaluated on $\tilde g_*$ due to \eqref{eq:1st}.  As a result, if we parametrize the metric in such a way that $\beta$ is linear in our coordinates $\tilde g$, then positivity of the second order term in $-A/4G$ under fluctuations that preserve $H$ is also equivalent to positivity of the 2nd order term in $I$ under fluctuations that preserve $A$.  As we describe in section \ref{sec:SL}, this leads to a natural Sturm-Liouville-type problem.


\section{Formulation of the stability problem}
\label{sec:SL}

The strategy we will follow in studying stability of our microcanonical saddles will be essentially identical to that used for the canonical ensemble in \cite{CanoPaper}.  We review this approach briefly below. Since the purpose of this section is to make contact with our later explicit calculations, we now change notation to again parallel traditional choices made in analyzing the existence of such negative modes.  In particular, in this section we will write all indices explicitly and drop the\ $\tilde{}$\ over quantities associated with smooth metrics.

From a conceptual point of view, the most important issue to discuss is to fix a contour of integration for the path integral.  After all, given a saddle in the complex plane, at least locally one can always find a steepest descent contour along which fluctuations about the saddle would be stable.  On the other hand, one can similarly always find a steepest {\it ascent} contour along which fluctuations about the same saddle would be unstable.  So an investigation of stability can have meaning only if we first fix the contour of integration.

Recall that a purely-real Euclidean contour is not viable (see e.g. \cite{Gibbons:1978ac}).  Due to the conformal factor problem, such a choice would imply that {\it all} saddles are badly unstable.  A traditional choice has been to Wick-rotate pure-trace modes \cite{Gibbons:1978ac} or some direct analogue 
\cite{Monteiro:2008wr} 
 (perhaps found by following the approach of \cite{Kol:2006ga}) while preserving the reality of the Euclidean transverse-traceless modes.  See also \cite{Gratton:1999ya,Gratton:2000fj,Gratton:2001gw} in the cosmological context. But our microcanonical boundary conditions require us to fix the area of the Euclidean horizon, which will turn out to couple the pure-trace and transverse-traceless modes. It is thus impossible to Wick-rotate the former without also Wick-rotating the latter.

This, however, is precisely the same issue that arises in the study of the canonical ensemble in a finite cavity \cite{CanoPaper}.  We may thus use the rule-of-thumb described in that work  to fix the contour of integration.  Though we refer the reader to the original reference for details, the end result is determined by the spectrum of an operator ${\mathbb L}$ defined by first choosing an inner product on the space of perturbations $h_{ab}$ and then combining this metric with second functional derivatives of the action.    In particular, suppose we have an action $\check{\mathbb{I}}$ (the reason for the check ($\check{}$) will be explained shortly).  Suppose also that $\check{\mathbb{I}}$ is a quadratic function of independent and unconstrained\footnote{In particular, the $\mathbb{Q}^I$ represent the parts of the fields that remain free after imposing any boundary conditions.  The boundary conditions are then incorporated into the definition of $\check{\mathbb {I}}$ as in \cite{CanoPaper}.} real field variables ${\mathbb{Q}}^I$ that is stationary at $\mathbb{Q}^I=0$.  If we also choose a real inner product
\begin{equation}
\label{eq:IP}
(\mathbb{Q}, \mathbb{Q}') = \sum_{IJ}    \mathbb{Q}^I\mathbb{Q}^J \mathbb{G}_{IJ}, 
\end{equation}
on the space of unconstrained variations $Q^I$, the relevant operator $\mathbb{L}$ is defined by 
\begin{equation}
\label{eq:linop}
{\mathbb{L}}_J^I = \sum_K {\mathbb{G}}^{IK} \check{\mathbb{S}}_{,KJ}.
\end{equation}
so that the action is the expectation value of $\mathbb{L}$ in \eqref{eq:IP}. 
In this general discussion the label $I$ may take either continuous values (as in the problem we wish to study) or discrete values (as in the numerical approximations that we will use in practice).  In the latter case $\sum_I, \sum_{IJ}$ represents the appropriate integral and $,IJ$ denotes functional derivatives. 

The operator $\mathbb{L}$ is always self adjoint with respect to the inner product \eqref{eq:IP}.
It is assumed to be diagonalizeable, though, since the inner product we choose will not be positive definite, the spectrum of ${\mathbb L}$ is generally complex.  After the Wick rotation, a mode with eigenvalue $\lambda$ is stable if and only if ${\rm Re} \ \lambda > 0$.

As in \cite{CanoPaper}, we use the DeWitt$_{-1}$ inner product. In the continuum (and reverting to more standard continuum notation) this is given by
(and reverting to more standard continuum notation) 
\begin{equation}
(h,h') = 
\frac{1}{32 \pi G}\int_{\mathcal{M}}\mathrm{d}^d x\,\sqrt{\hat{g}}\,h_{ab}\,\hat{\mathcal{G}}^{ab\;cd} h_{ab}{}',
\label{eq:DWIP}
\end{equation}
with
\begin{equation}
\label{eq:cDWIP}
 \hat{\mathcal{G}}^{ab\;cd}=\frac{1}{2}\left(\hat{g}^{ac}\hat{g}^{bd}+\hat{g}^{ad}\hat{g}^{bc}-\hat{g}^{ab}\hat{g}^{cd}\right)\,.
\end{equation}
The $-1$ in the subscript on DeWitt$_{-1}$ refers to the coefficient of the final term in \eqref{eq:cDWIP}.
Due to this choice, $\mathbb{L}$ will again be just the familiar \L operator in the bulk, though it will be equipped with new boundary conditions and, in addition, the operator itself will now contain an additional boundary term at the cavity wall.

To explain this point in detail, we expand a general metric $g_{ab}$ near a saddle $\hat{g}_{ab}$ as
\begin{equation}
g_{ab}=\hat{g}_{ab}+\varepsilon\,h_{ab},
\end{equation}
where $\varepsilon$ is a bookkeeping parameter that we take to be infinitesimally small. As above, let $I$ denote the Euclidean Einstein-Hilbert action with a Gibbons-Hawking-York term at the cavity wall.  To second order in $\varepsilon$, this action may be written \cite{Headrick:2006ti} 
\begin{subequations}
\begin{equation}
\label{eq:I(2)}
I^{(2)} = \frac{\varepsilon^2}{32 \pi G}\int_{\mathcal{M}}\mathrm{d}^d x\,\sqrt{\hat{g}}\,h_{ab}\,\hat{\mathcal{G}}^{ab\;cd}\left[(\hat{\Delta}_L h)_{cd}+2 \hat{\nabla}_{(c}\hat{\nabla}^p \bar{h}_{d)p}\right]+\delta I^{(2)}_{\partial}
\end{equation}
where
\begin{equation}
\bar{h}_{ab}=h_{ab}-\frac{\hat{g}_{ab}}{2}h\,,
\end{equation}
\begin{equation}
(\hat{\Delta}_L h)_{ab}=-\hat{\nabla}_p \hat{\nabla}^p h_{ab}-2\,\hat{R}_{acbd}\,h^{cd}\,,
\end{equation}
\begin{equation}
 \hat{\mathcal{G}}^{ab\;cd}=\frac{1}{2}\left(\hat{g}^{ac}\hat{g}^{bd}+\hat{g}^{ad}\hat{g}^{bc}-\hat{g}^{ab}\hat{g}^{cd}\right)\,.
\end{equation}
\end{subequations}%
Here $\delta I^{(2)}_{\partial}$ is a boundary term at the cavity wall 
whose detailed form can be found in the appendix \ref{app:1}.

Now, if one imposes boundary conditions appropriate to the canonical ensemble, the boundary term $\delta I^{(2)}_{\partial}$ vanishes.  It is possible that we might achieve a similar result in the microcanonical case by working directly with the microcanonical action.  But to make contact with familiar technology, we instead use the  strategy described at the end of section \ref{sec:sandf} that involves working with $I$, but also implementing appropriate microcanonical boundary conditions.  

Recall that these boundary conditions enforce two conditions at the Euclidean horizon:  lack of a conical singularity and preservation of the horizon area.  In fact, for static modes with zero angular momentum these are the {\it only} boundary conditions required to be enforced\footnote{Though, in dealing with the gauge symmetries, we will shortly choose to impose both the de Donder gauge condition  and its derivative there as well.}.  In particular, for such modes there are no boundary conditions at the cavity wall.  As a result, the boundary conditions are of little use in simplifying $\delta I^{(2)}_{\partial}$.  This boundary term turns out to be both complicated and subtle.

We find it useful to deal with such subtleties by working directly with a discretized version of the system and then numerically investigating the continuum limit.  This is in accord with the general strategy described in section 2.2  of \cite{CanoPaper}.  After discretizing the action and expressing the result in terms of fields not constrained by boundary conditions, it is straightforward to take derivatives of the action and then to define $\mathbb{L}$ via \eqref{eq:linop}.  This contrasts with the traditional approach of \cite{Gibbons:1978ji,Gross:1982cv,Allen:1984bp,Prestidge:1999uq,Headrick:2006ti,Monteiro:2009ke,Monteiro:2009tc,Dias:2009iu,Dias:2010eu} which used the continuum action to define a continuum differential operator, after which the boundary conditions could be used to define an appropriate discretization of this operator.  

The one remaining issue to be discussed involves the diffeomorphism gauge symmetries.  The main point here is that each such symmetry gives rise to zero-eigenvalue eigenvectors of ${\mathbb L}$, making the spectrum of ${\mathbb L}$ highly degenerate.  This is technically troublesome, so we will take steps to remove this degeneracy.

Here, again, we follow the treatment of \cite{CanoPaper} and we refer the reader to that reference for details (see in particular its appendix A).  
In essence, we will impose the de Donder gauge condition
\begin{equation}
    \label{eq:DDGC}
    \nabla^a h_{ab} - \frac{1}{2} \nabla_b h  =0.
\end{equation} For static spherically-symmetric perturbations about Euclidean AdS-Schwarzschild, the time-component of the de Donder gauge condition is straightforward to implement.    Indeed, doing so corresponds merely to setting the $\tau r$ component of the metric perturbation to zero, where $r$ is the usual area-radius and $\tau$ is a standard Killing time coordinate on AdS-Schwarzschild.  However, it is non-trivial to impose the remaining radial component.   While it can also be solved, inserting the solution into the action leads to a higher-derivative action that is more complicated to study.

As a result, we will not solve the radial component of the de Donder condition explicitly (though it will be convenient to impose both the condition and its derivative precisely at the cavity wall as a extra boundary conditions).  Instead, we will call the quadratic gravitational action ${\mathbb{I}}$ (with no check ($\check{}$)), and we will define a new action 
\begin{equation}
\label{eq:Smod}
\check{\mathbb{I}}(\mathbb{Q}) \equiv {\mathbb{I}}(\mathbb{Q}) + \sum_{IJ} \mathbb{K}_{IJ} \mathbb{Q}^I\mathbb{Q}^J
\end{equation}
 to be used in our numerics by adding the additional quadratic term $\sum_{IJ} \mathbb{K}_{IJ} \mathbb{Q}^I\mathbb{Q}^J$ that breaks gauge invariance\footnote{Simple discretizations of an action generally break exact gauge invariance in any case.  But when it results from discretization, this gives only a tiny splitting between the eigenvalues of $\mathbb{L}$ associated with would-be gauge degrees of freedom, and in particular this splitting disappears when the discretization scale is taken to zero so that the continuum limit is restored.  We find it useful create a larger splitting by adding by hand an additional term that breaks gauge invariance and whose effects do not vanish in the continuum limit.} but which has no effect on modes that satisfy the de Donder condition.  That is to say, we require $\sum_{J} \mathbb{K}_{IJ} \mathbb{Q}^J =0$ when $\mathbb{Q}^I$ satisfies the de Donder condition. As in \cite{CanoPaper}, we choose this additional term to precisely cancel the $2 \hat{\nabla}_{(c}\hat{\nabla}^p \bar{h}_{d)p}$ term in \eqref{eq:I(2)}.  As a result, in the bulk the resulting $\mathbb{L}$ will just be the usual \L operator, or a discretized version thereof, though it is now supplemented by novel boundary terms and boundary conditions.  As explained in \cite{CanoPaper}, the properties of the de Donder gauge then imply that the eigenmodes of $\mathbb{L}$ will sort themselves nicely into modes satisfying the de Donder condition and modes that are pure gauge. 
With these details in hand, we are now ready to analyze our gravitational system.

\section{The Euclidean Schwarzschild-AdS black hole and perturbations thereof}
\label{sec:perts}

In this work we will focus on studying the stability properties of the Euclidean Schwarzschild black hole, whose line element can be written as
\begin{subequations}
\begin{equation}
\mathrm{d}s^2=f(r)\mathrm{d}\tau^2+\frac{\mathrm{d}r^2}{f(r)}+r^2 \mathrm{d}\Omega_{d-2}^2\,,
\end{equation}
where recall that $\mathrm{d}\Omega_{d-2}^2$ is the line element on a round $S^{d-2}$ sphere with unit radius,
\begin{equation}
f(r)=\frac{r^2}{\ell^2}+1-\left(\frac{r_+}{r}\right)^{d-3}\left(\frac{r_+^2}{\ell^2}+1\right)
\end{equation}
\end{subequations}%
$\ell$ is the AdS radius and $r=r_+$ is the location of the black hole horizon, so that $f(r_+)=0$.

We shall be interested in metric perturbations that preserve the spherical background symmetry. Since the microcanonical path integral of section \ref{sec:gen} was constructed from the canonical path integral by performing only a {\it single} extra integral over the zero angular-momentum mode described by $T$, perturbations that carry momentum on the $S^{d-2}$ sphere will satisfy the same boundary conditions as for the canonical ensemble.  Furthermore, in the infinite cavity limit, adding angular momentum has been shown in \cite{Kudoh:2006bp} to increase the eigenvalues of the {\L} operator. We will assume that the corresponding result continues to hold in the presence of a finite-radius spherical wall and reserve a careful check of this assumption for future work.

Since the background metric is spherically symmetric, we can use the background symmetry to decompose metric perturbations into tensor, vector and scalar derived perturbations. These are metric perturbations built from tensor, vector and scalar harmonics, respectively. Since tensor and vector harmonics necessarily carry momentum, they are excluded using our previous argument. We are thus left with a spherically symmetric scalar perturbation.  Imposing the $\tau$ component of the de Donder condition \eqref{eq:DDGC} sets the $\tau r$ component of the perturbation to zero, leaving us with
\begin{equation}
\delta \mathrm{d}s^2 = a(r)f(r)\mathrm{d}\tau^2+\frac{b(r)}{f(r)}\mathrm{d}r^2+c(r)\,r^2 \mathrm{d}\Omega_{d-2}^2\, .
\label{eq:lineSO3}
\end{equation}
The perturbed line element is then a function of three unknown functions of $r$: $\{a,\,b,\,c\}$.

At this point we introduce a spherical cavity at the coordinate location $r=r_0$. At the walls of this cavity we demand that the sphere sphere radius is not changing, but leave the period $\beta$ free to vary.

Evaluated on the perturbed line element (\ref{eq:lineSO3}), the modified second order action defined at the end of section \ref{sec:sandf} reads \cite{CanoPaper}
\begin{multline}
{I}^{(2)} =\frac{\varepsilon^2 \Omega_{d-2}}{64 \pi G}\Bigg\{\int_{r_+}^{r_0} \mathrm{d}r\,r^{d-2}\vec{q}\cdot\left[\frac{1}{r^{d-2}}\mathbf{P}\cdot \frac{\mathrm{d}}{\mathrm{d}r} \left(f\,r^{d-2} \frac{\mathrm{d}\vec{q}}{\mathrm{d}r}\right)+\mathbf{V}\cdot \vec{q}\right]
\\
+2(d-2)r_0^{d-3}\left[a(r_0)+b(r_0)-r_0 c^\prime(r_0)\right]\Bigg\}\,,
\label{eq:actionneg}
\end{multline}
where $\Omega_{d-2}$ is the volume of the metric on a unit radius round $(d-2)-$sphere, $\vec{q}=\{a(r),b(r),c(r)\}$, the operation $\cdot$ denotes the standard Cartesian inner product in Euclidean space defined by the Kronecker delta metric $\delta_{\tilde I, \tilde J}$, and $\mathbf{P}$ and $\mathbf{V}$ are symmetric matrices with the following independent components
\begin{equation}
\mathbf{P}_{11}=\mathbf{P}_{22}=-1\,, \quad \mathbf{P}_{12}=1\,,\quad  \mathbf{P}_{13}=\mathbf{P}_{23}=d-2\,,\quad\text{and}\quad  \mathbf{P}_{33}=(d-4)(d-2)\, ,
\end{equation}
and
\begin{align}
&\mathbf{V}_{11}=-\frac{(d-2) f'(r)}{r}+\frac{f'(r)^2}{f(r)}-f''(r) \nonumber
\\
& \mathbf{V}_{12}=-\frac{(d-2) f'(r)}{r}-\frac{f'(r)^2}{f(r)}+f''(r)\nonumber
\\
& \mathbf{V}_{13}=\frac{(d-3) (d-2)}{r^2}-\frac{(d-3) (d-2) f(r)}{r^2}-\frac{(d-4) (d-2) f'(r)}{2 r}-\frac{1}{2} (d-2) f''(r)\nonumber
\\
& \mathbf{V}_{22}=\frac{4 (d-2) f(r)}{r^2}-\frac{(d-2) f'(r)}{r}+\frac{f'(r)^2}{f(r)}-f''(r)\nonumber
\\
& \mathbf{V}_{23}=\frac{(d-3) (d-2)}{r^2}-\frac{(d-2) (d+1) f(r)}{r^2}-\frac{(d-4) (d-2) f'(r)}{2
   r}-\frac{1}{2} (d-2) f''(r)\nonumber
\\
& \mathbf{V}_{33}=\frac{2 (d-4) (d-3) (d-2)}{r^2}-\frac{2 (d-5) (d-2)^2 f(r)}{r^2}-\frac{2 (d-2)^2 f'(r)}{r}\,.
\end{align}

Because our problem has reduced co-homogeneity, the microcanonical boundary conditions are trivial to state.  At the horizon we require
\begin{equation}
c(r_+)=0,
\end{equation}
while $a$ and $b$ should be smooth at $r_+$. At first it might appear that our are insufficient at $r=r_+$. In fact, this is far from true. In particular, because $r=r_+$ is a regular singular point, smoothness of $a$, $b$ and $c$ at the horizon alone is enough to provide Robin boundary conditions for both $b$ and $c$ and to enforce $a(r_+)=b(r_+)$.  It might then appear that further imposing the condition $c(r_+)=0$ would lead to too many boundary conditions. However, at $r=r_0$ we only require that $c(r_0)=0$, while allowing $a(r_0)$ (which effectively controls $\beta$) to be arbitrary. 

As a result, we see that we in fact have five boundary conditions. This is not yet enough boundary conditions to make our problem well-posed. Perhaps more explicitly, this is not yet enough boundary conditions to allow us to define a consistent second-order differencing scheme that gives rise to a well-defined discretized version of the action.  We also see that it would be natural to expect an extra boundary condition at $r_0$.  To proceed, we simply impose the de Donder gauge condition \eqref{eq:DDGC} at the cavity wall, which then provides a Robin-type boundary condition for $b^{\prime}(r_0)$. We thus have precisely six boundary conditions in total, which is the correct number to define a discretized version of a second-order action for three fields.

We can of course also use the boundary conditions to define a discretized version $\mathbb G^{IJ}$ of the DeWitt$_{-1}$ metric.  However, this $\mathbb G^{IJ}$ turns out to be degenerate.  This differs from the situation with canonical boundary conditions discussed in \cite{CanoPaper}.  The issue appears to be related to additional gauge transformations that are allowed in the microcanonical context, as we find that this degeneracy can be removed by imposing an additional piece of the de Donder condition \eqref{eq:DDGC}.  Namely, we impose the derivative of \eqref{eq:DDGC} at the cavity wall:
    \begin{equation}
    \label{eq:DDDGC}
    \left(\nabla^b\nabla^a h_{ab} - \frac{1}{2} \nabla^2 h\right)|_{r=r_0}  =0.
\end{equation}
This condition can then be incorporated into the definition of the actions $\check{\mathbb{I}}$ and $\mathbb{I}$ as well.

\section{Results}
\label{sec:res}
We now report the results of our numerical investigations.  The action defined by combining the discussion of section \ref{sec:sandf} with the boundary conditions of section \ref{sec:perts} may be discretized using the method described in \cite{CanoPaper}. Combining this with the analogously discretized DeWitt$_{-1}$ metric defines an operator  $\mathbb{L}$ via \eqref{eq:linop}.  Since the system has already been discritized, this $\mathbb{L}$ is just a large matrix whose eigenvalues and eigenvectors can be studied using standard numerical methods \cite{Dias:2015nua}.

As in \cite{CanoPaper}, to decrease the clutter, we define the dimensionless quantities
\begin{equation}
\tilde{\lambda}\equiv \lambda\,r_+^2\,,\quad y_0\equiv \frac{r_0}{r_+}\,,\quad\text{and}\quad y_+\equiv \frac{r_+}{\ell}\,.
\end{equation}
The limit of zero cosmological constant is obtained by setting $y_+\to0$.
 
\begin{figure}[h]
\centering\includegraphics[width=.8\linewidth]{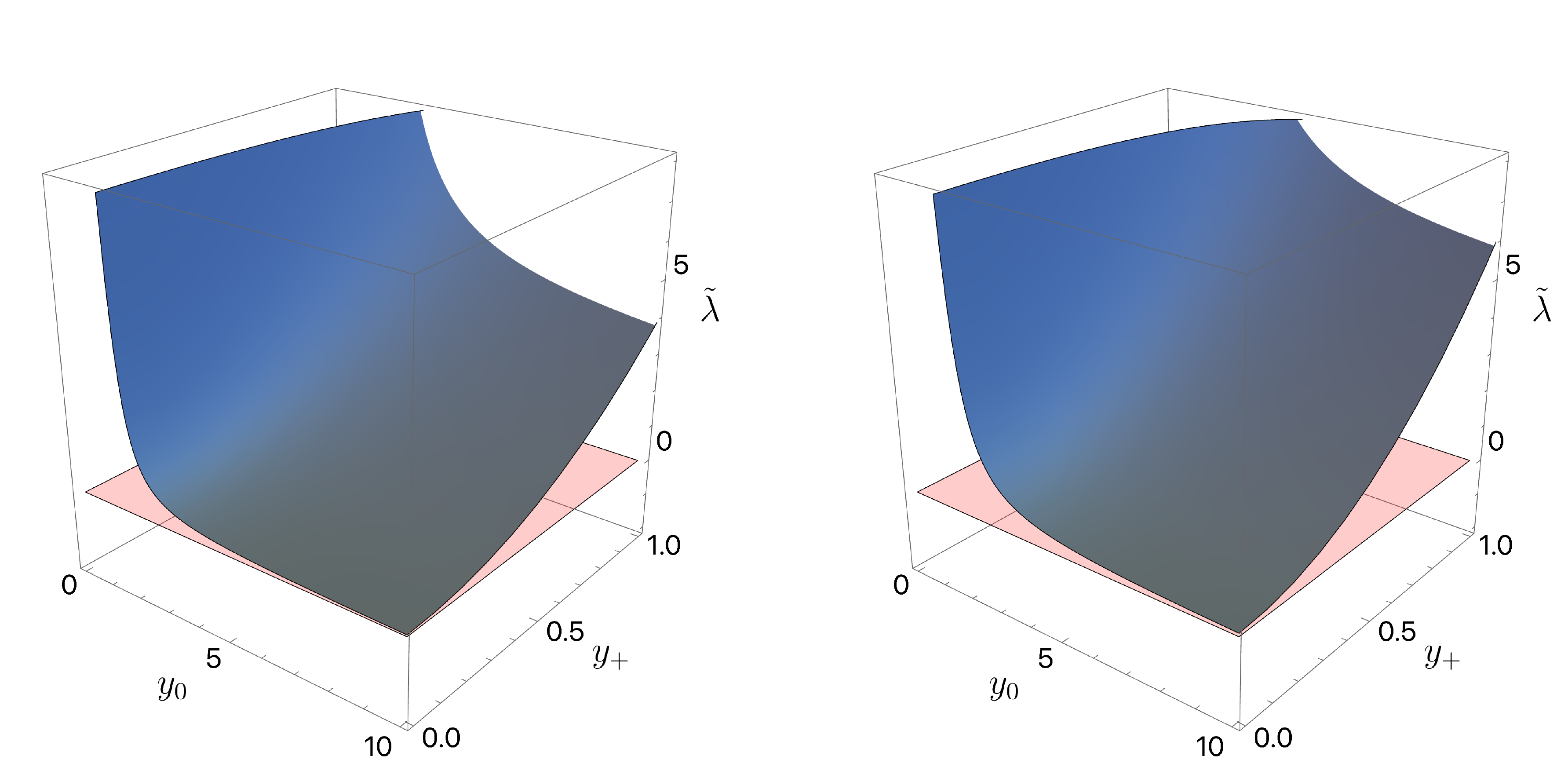}
\caption{The lowest lying eigenmode with microcanonical boundary conditions as a function of $y_0$ and $y_+$. On the left panel we have $d=4$, while on the right panel we take $d=5$. To aid visualisation we also plot the plane $\tilde{\lambda}=0$ in red. This mode turns out to be purely real and, most importantly, it has positive eigevanlue.}
\label{fig:pos}
\end{figure}

 Fig.~\ref{fig:pos} displays results for the lowest lying eigenmode (\emph{i.e.}, the mode with smallest $\rm{Re} \ \lambda$) as a function of $y_0$ and $y_+$ for $d=4$ (left panel) and $d=5$ (right panel). This mode turns out to be real and positive. This shows that the Euclidean Schwarzschild-AdS black hole is stable in the microcanonical ensemble, as expected. In addition to what is shown in the figure,  with $y_+=0$ we extended our results to $y_0\sim 40$ and found the lowest mode to always remain real.

The positive eigenvalue indicates that the mode displayed in Fig.~\ref{fig:pos} is stable after making the Wick rotation proposed in \cite{CanoPaper}.  But, as an aside, it is interesting to ask if this particular mode is one of the ones to be Wick-rotated, or it if is to be left along the real Euclidean axis.  The answer to that question is determined by studying the norm of the mode under \eqref{eq:IP}, as for real modes the rule-of-thumb from \cite{CanoPaper} rotates only those with negative norm. 
It turns out that our lowest-lying mode has positive norm, and thus that it defines a stable perturbation even before any Wick rotation. To show this, we note that any metric perturbation can be decomposed as a sum of a traceless component $\tilde{h}_{ab}$ and a pure trace part $\phi$:
\begin{equation}
h_{ab}=\tilde{h}_{ab}+\frac{1}{d}\hat{g}_{ab}\,\phi\,.
\end{equation}
The metric $\hat{\mathcal{G}}$ is such that $\tilde{h}_{ab}$ and $\hat{g}_{ab}\,h$ are orthogonal to each other. That is to say
\begin{equation}
\tilde{h}_{ab} \hat{G}^{ab\,cd}\hat{g}_{cd}=0\,.
\end{equation}
This, in turn, implies that the norm under $\hat{\mathcal{G}}$ may be written in the form
\begin{subequations}
\begin{align}
\lVert h \rVert^2& \equiv \frac{1}{32 \pi G}\int_{\mathcal{M}} \mathrm{d}^d x\sqrt{\hat{g}}\;h_{ab}\hat{\mathcal{G}}^{ab\,cd}h_{cd}\nonumber
\\
& = \frac{1}{32 \pi G}\left(\int_{\mathcal{M}} \mathrm{d}^d x\sqrt{\hat{g}}\;\tilde{h}_{ab}\;\tilde{h}^{ab}- \int_{\mathcal{M}} \mathrm{d}^d x\sqrt{\hat{g}}\;\phi^2\right)\nonumber
\\
& \equiv\lVert \tilde{h} \rVert^2_{\infty}-\lVert \phi \rVert^2_{\infty}\,,
\end{align}
where we have defined
\begin{equation}
\lVert \tilde{h} \rVert^2_{\infty}\equiv \frac{1}{32 \pi G}\int_{\mathcal{M}} \mathrm{d}^d x\sqrt{\hat{g}}\;\tilde{h}_{ab}\;\tilde{h}^{ab}>0
\end{equation}
and
\begin{equation}
\lVert \phi \rVert^2_{\infty}\equiv \frac{1}{32 \pi G}\frac{d-2}{2d}\int_{\mathcal{M}} \mathrm{d}^d x\sqrt{\hat{g}}\;\phi^2>0\,.
\end{equation}
\end{subequations}
It thus follows that positivity of $\lVert h \rVert^2$ is equivalent to positivity of 
\begin{equation}
\eta\equiv 1-\frac{\lVert \phi \rVert^2_{\infty}}{\lVert \tilde{h} \rVert^2_{\infty}}\, .
\label{eq:eta}
\end{equation}

\begin{figure}[h]
\centering\includegraphics[width=.8\linewidth]{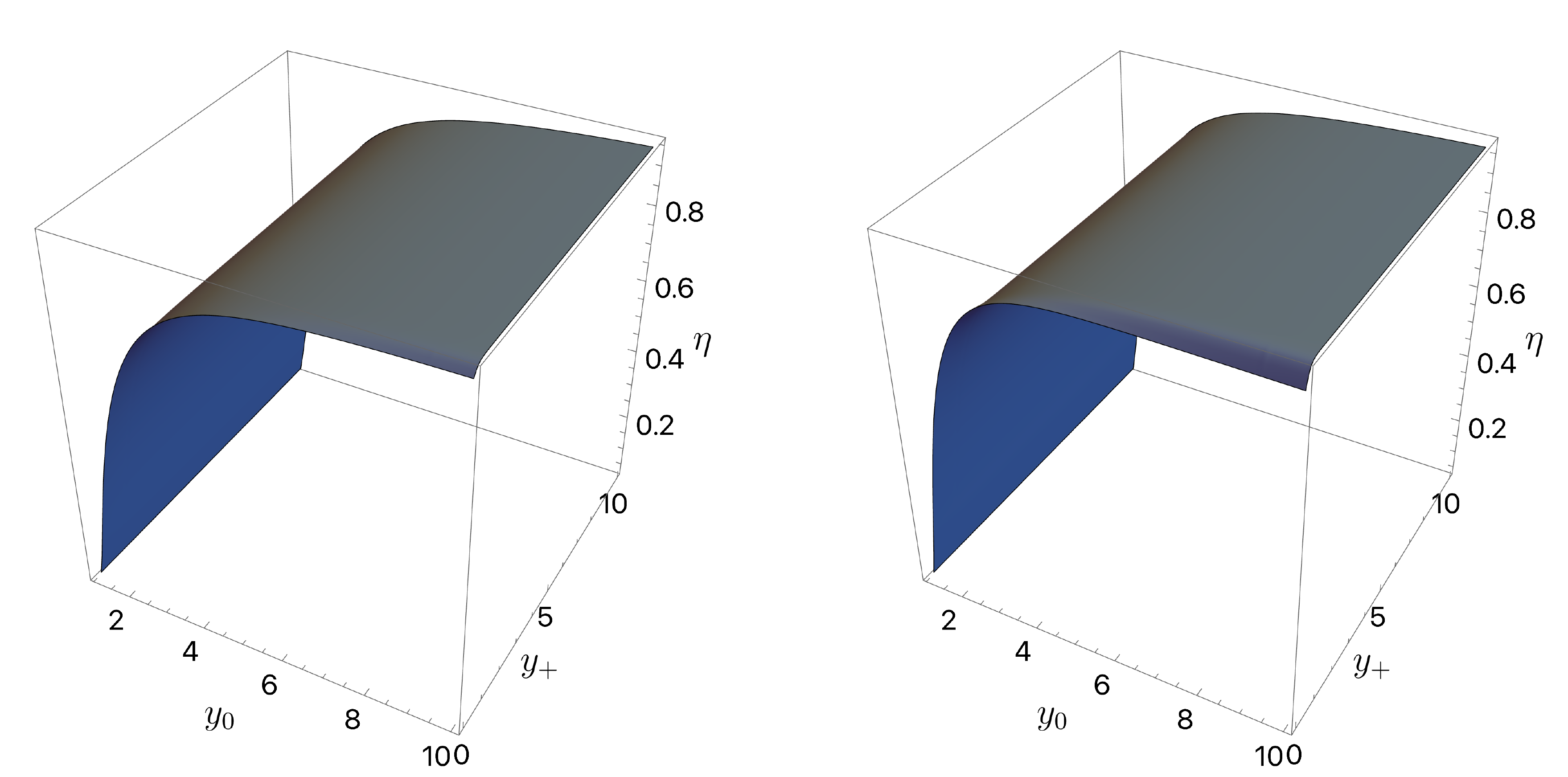}
\caption{The quantity $\eta$ computed for the lowest lying eigenmode with microcanonical boundary conditions as a function of $y_0$ and $y_+$. On the left panel we have $d=4$, while on the right panel we take $d=5$.  In both cases the results indicate that the mode has positive norm.  As a result, the rule-of-thumb from \cite{CanoPaper} does not Wick-rotate either mode.  And since we saw above that they have positive eigenvalues, both modes also define stable directions of the original action before any Wick-rotation.}
\label{fig:norm}
\end{figure}

In Fig.~\ref{fig:norm} we plot $\eta$ as a function of $y_0$ and $y_+$ and find that it is positive. This confirms that the lowest lying mode has positive norm under $\hat{\mathcal{G}}$.  When combined with the positive eigenvalue found above, it also shows that the mode defines a stable perturbation even before Wick-rotating the contour of integration.  This completes the aside mentioned above.
\begin{figure}[h]
\centering\includegraphics[width=.8\linewidth]{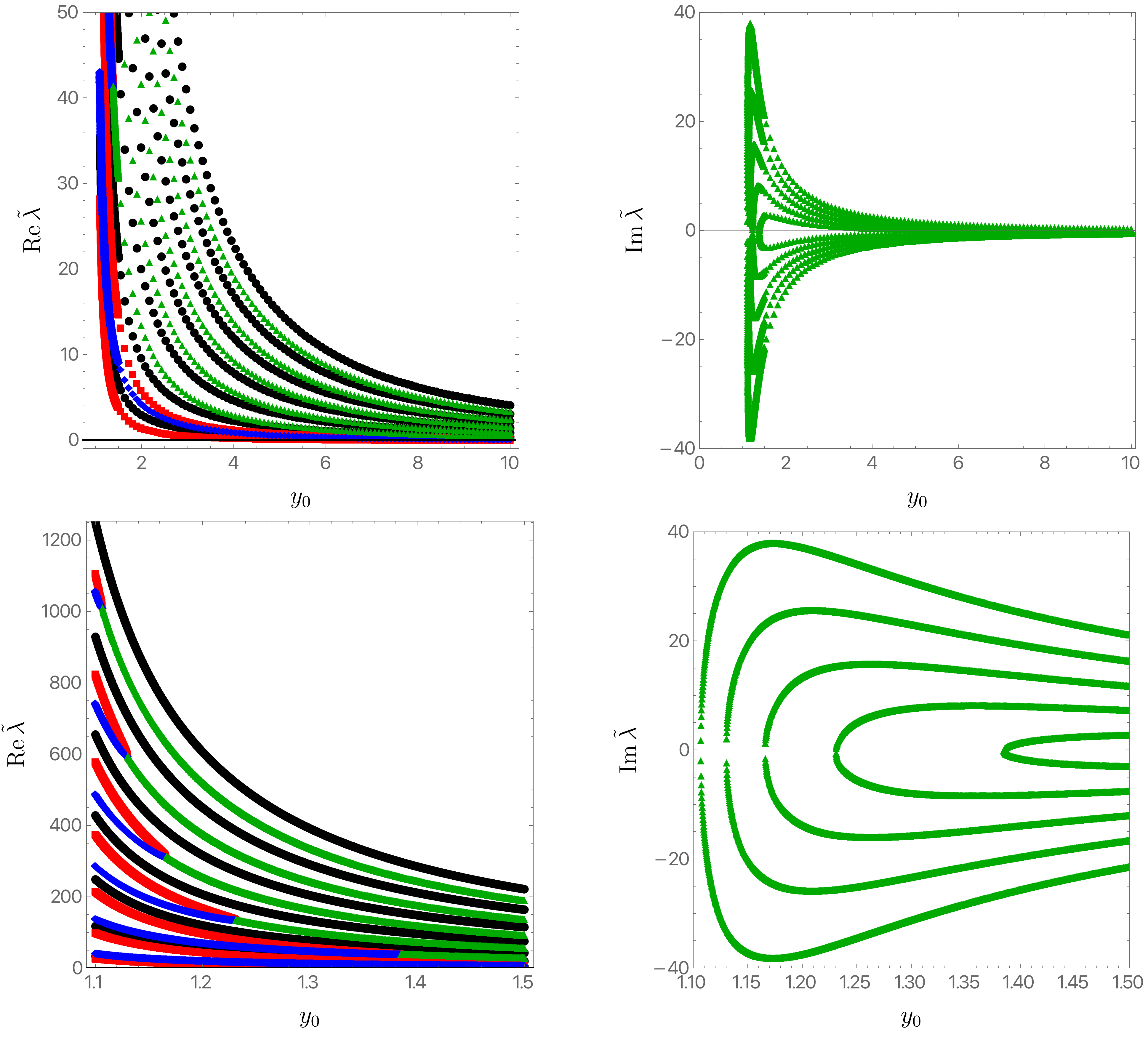}
\caption{The real part (left column) and the imaginary part (right column) of the excited modes as a function of $y_0$ for $y_+=0$. The figures in the bottom row show magnified versions of small regions from the figures in the top row.  All plots are for $d=4$. The colour coding is as follows: green triangles are non-gauge modes with complex eigenvalues; blue diamonds are non-gauge modes with negative norm under $\hat{\mathcal{G}}$; red squares are non-gauge modes with positive norm under $\hat{\mathcal{G}}$ and the black disks are pure gauge modes. Each green triangle has a two-fold degeneracy since complex modes come in conjugate pairs.}
\label{fig:4D}
\end{figure}

We now attempt to give a more comprehensive survey of the overtones. For concreteness, we take $y_+=0$, but we find similar results for $y_+\neq0$. In general, the eigenvalues $\lambda$ are complex, just as with canonical boundary conditions \cite{CanoPaper}. However, for the canonical boundary conditions discussed in \cite{CanoPaper}, at any given $y_0, y_+$ only a small number of the  eigenvalues were $\lambda$ complex.  In contrast, when considering microcanonical boundary conditions, at least when the cavity is large enough we find almost all eigenvalues to be complex. In Fig.~\ref{fig:4D} we plot the first twenty overtones in $d=4$ as a function of $y_0\in(1,10)$ for $y_+=0$.
The bottom row shows magnified versions of small regions from the figures in the top row.  The left column shows the real part of the eigenvalue and the right column shows the corresponding imaginary part. 
Just as for canonical boundary conditions, we see that a negative and positive norm modes can merge to form a complex mode.
 Due to the approach described in section \ref{sec:sandf}, some fraction of the modes shown in figure \ref{fig:4D} are pure gauge.  Modes are classified as gauge vs. non-gauge by comparing the modified action $\check{S}{}^{(2)}$ to the original action ${S}{}^{(2)}$, with vanishing of the latter (to numerical precision) indicating a gauge-mode while for non-gauge modes the values agree. As a consistency check, we also verify that the non-gauge modes are precisely those that satisfy the de Donder gauge condition as expected.
\begin{figure}[h]
\centering\includegraphics[width=.8\linewidth]{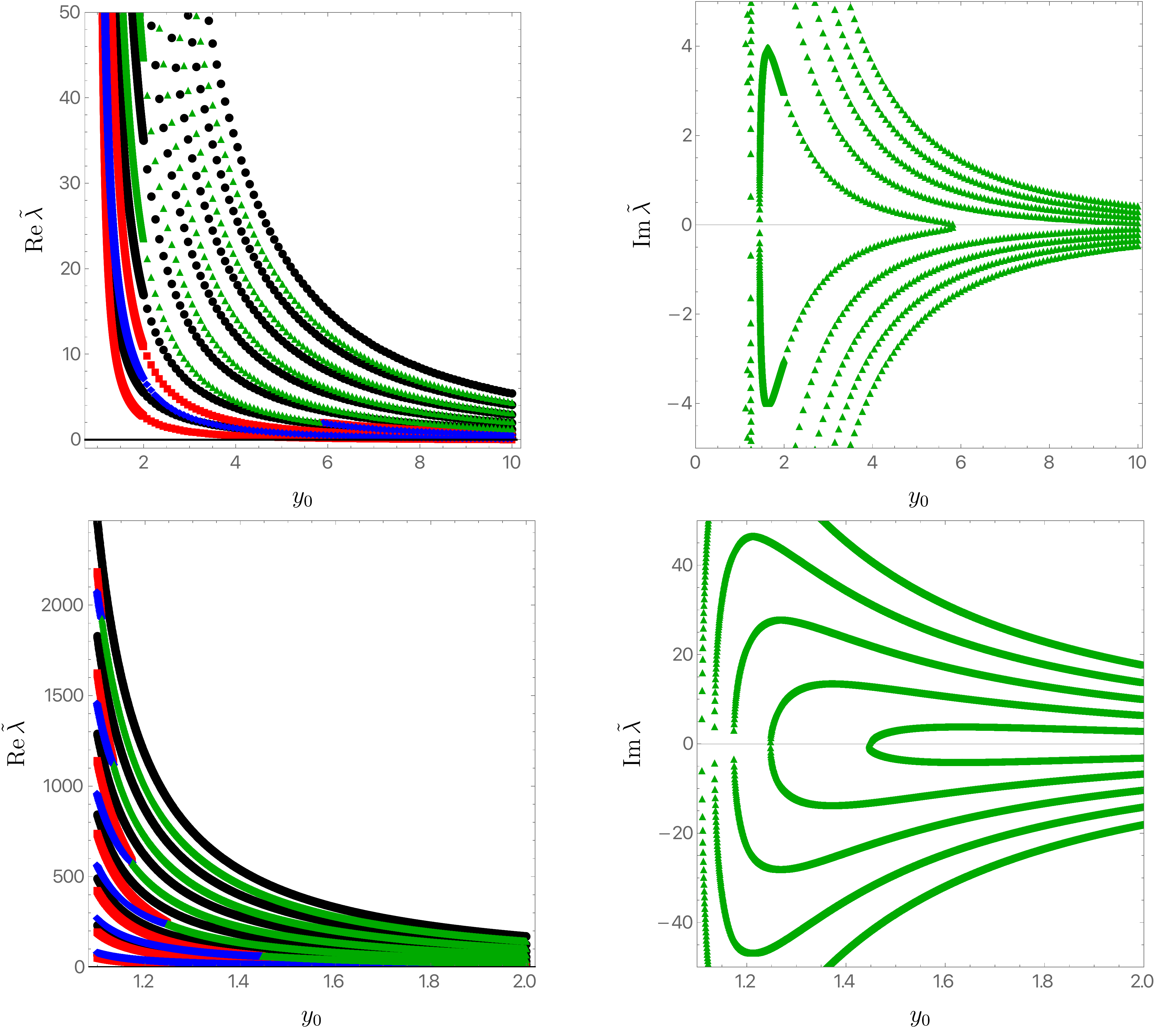}
\caption{The real part (left column) and the imaginary part (right column) of the excited modes as a function of $y_0$ for $y_+=0$. The bottom row is a zoom of the top row and all plots have $d=5$. The colour coding is as follows: green triangles are non-gauge modes with complex eigenvalues; blue diamonds are non-gauge modes with negative norm under $\hat{\mathcal{G}}$; red squares are non-gauge modes with positive norm under $\hat{\mathcal{G}}$ and the black disks are pure gauge modes. Each green triangle has a two-fold degeneracy since complex modes come in conjugate pairs.}
\label{fig:5D}
\end{figure}

We have checked that the rule of thumb presented in \cite{CanoPaper} remains well-defined even when the positive and negative norm modes merge. In particular, using the normalisation and phase conventions of \cite{CanoPaper} for the complex modes, we find that the imaginary part of any complex modes becomes precisely the negative norm mode after the merger. Similarly, the real part of any complex mode becomes the corresponding positive norm mode after the merger.
 A similar structure can be seen in Fig.~\ref{fig:5D} which displays analogous results for $d=5$.

\begin{figure}[h]
\centering\includegraphics[width=.8\linewidth]{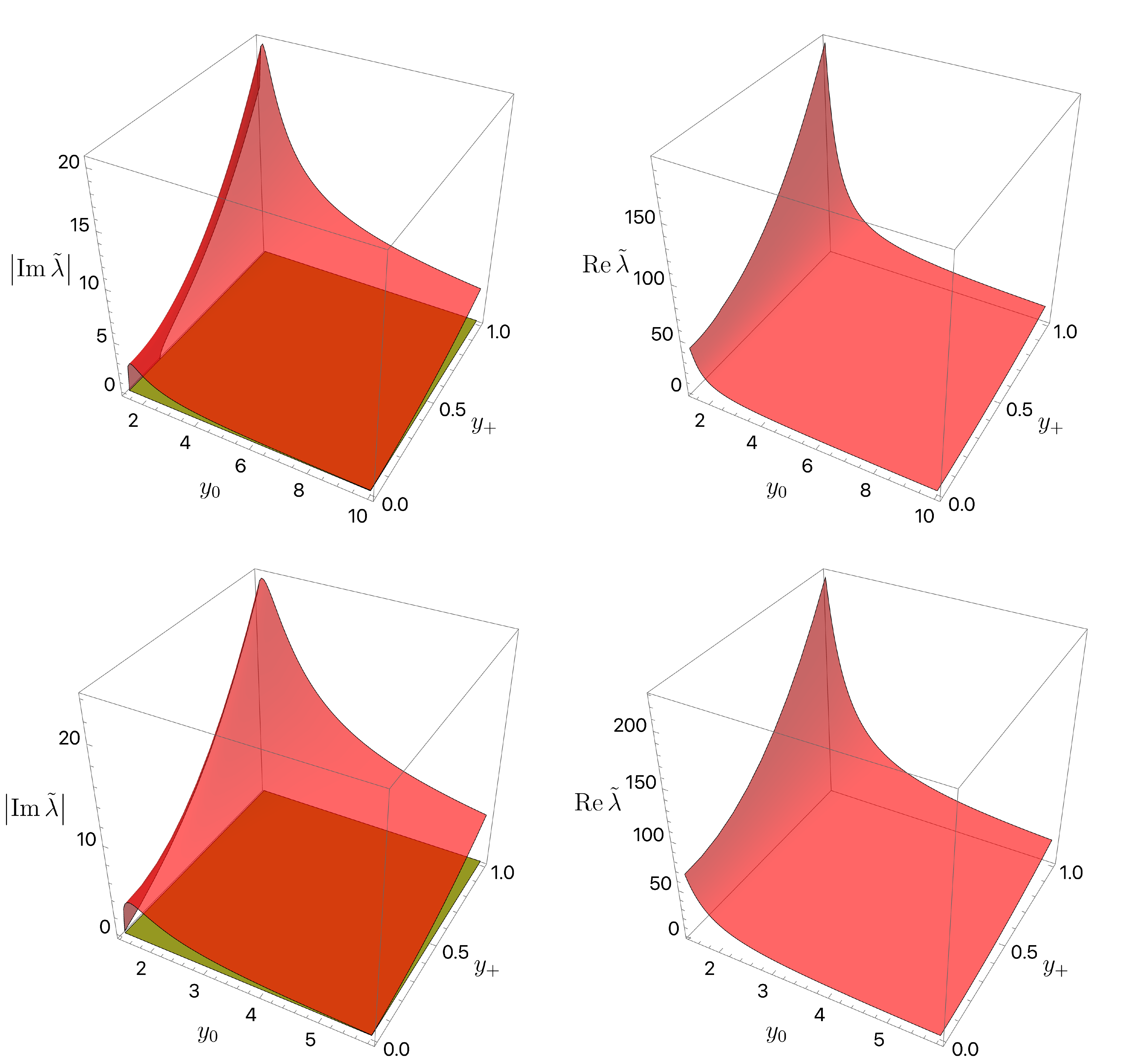}
\caption{The absolute value of the imaginary part (left column) and the real part (right column) of the first complex excited mode in a region of $y_0$ and $y_+$ where the mode is complex. Note that the scales are very different in the left and right figures.  The top row has $d=4$, while the bottom row has $d=5$. The green horizontal plane on the left column sits at $\tilde{\lambda}=0$ and is drawn solely to aid visualization.}
\label{fig:2dcomplex}
\end{figure}

Let us now generalize the discussion to 
$y_+ >0$ (\emph{i.e.}, to $\Lambda < 0$).
Corresponding results for the overtones in this case are shown in Fig.~\ref{fig:2dcomplex}.  There we plot the imaginary (left column) and real (right columns) of the first excited complex mode for $d=4$ (top row) and $d=5$ (bottom row). Unlike the case for the canonical boundary conditions studied in \cite{CanoPaper}, we now find that the size of the interval in $y_0$ where a given mode is complex now has only a weak dependence on $y_+$. In particular, we find large regions of parameter space where complex modes exist even for very large black holes (as large as $y_+=10$). 

\section{Discussion \& Conclusions}
\label{sec:disc}

The above work introduced a microcanonical path integral for gravitational systems that fixes a certain off-shell notion of energy.  The formulation was motivated by following \cite{Marolf:2018ldl} in writing the microcanonical partition function as an integral over the inverse temperature $\beta$ of the canonical partition, and then performing this $\beta$ integral before path-integrating over `the rest of the spacetime metric'.  This procedure is not unique, but depends on what aspects of the metric are held fixed during the $\beta$-integration.  

We simplify the problem by considering only metrics that preserve a $U(1)$ Euclidean time-translation symmetry that fixed a locus we may call the Euclidean horizon. 
In that context, for a natural choice of what it means to `fix the rest of the metric', we obtain a microcanonical path integral that specifies the value of a certain off-shell notion of the Hamiltonian that we call $H$. This $H$ is defined by including terms that would be called constraints in the canonical formalism, multiplied by appropriate notions of lapse and shift.  With this choice, and using the  Wick-rotation specified by the rule-of-thumb from \cite{CanoPaper},  we find  no microcanonical negative modes in perturbation theory about any Schwarzschild-AdS$_d$ black holes inside a cavity of any radius in pure Einstein-Hilbert asymptotically-AdS$_d$ gravity with $\Lambda \le 0$ for either $d=4$ or $d=5$.  In particular, including the constraint terms resolves the tension mentioned in the introduction that the negative mode in the asymptotically-flat canonical ensemble falls off too quickly to contribute to the usual ADM energy.
We presume the corresponding solutions to give stable microcanonical saddles in other dimensions as well.  While we worked with vacuum gravity, the general arguments did not depend on the detailed form of the action, so it is straightforward to extend the present formalism to include couplings to matter fields.

There are, however, a number of directions that remain to be explored.  First, one would like to study more interesting ensembles, such as those associated with rotating black holes.  This will be the subject of future work, where we anticipate finding that their study is facilitated by using the Hamiltonian form of the path integral.  Indeed, as indicated in footnote \ref{foot:measure}, the Hamiltonian formalism will also give better control over issues related to the path integral measure. We also expect the Hamiltonian formalism to allow the inclusion of modes that break the Euclidean time-translation symmetry, in which case we expect the corresponding method to fix a time-averaged notion of off-shell energy.

This then raises the question of studying negative modes in the Hamiltonian framework, where off-shell modes can involve fluctuations of canonical momenta that are not determined by time-derivatives of the spatial metric.  To our knowledge, this problem has not yet been studied in the literature.

Another important open question involves the choice of what it means to integrate over $\beta$ while holding the rest of the spacetime metric fixed.  Here we studied a particular  simple and perhaps natural such definition. One would expect the physics to be independent of such choices, but verifying this should be a high priority. 

We should also mention that there are cases where we {\it do} expect negative modes in the microcanonical path integral. In particular, we expect these to occur whenever the background solution is classically unstable. One such example occurs when we consider embedding the Euclidean Schwarzschild-AdS$_5$ in IIB supergravity with AdS$_5\times S^5$ asymptotics via the standard Freud-Rubin compactifications \cite{Freund:1980xh}. At small energies the usual Schwarzschild-AdS$_5$ black hole becomes dynamically unstable \cite{Martinec:strings98,Banks:1998dd,Peet:1998cr,Prestidge:1999uq,Hubeny:2002xn,Dias:2015pda,Buchel:2015gxa,Dias:2016eto,Cardona:2020unx} to the Gregory-Laflamme instability \cite{Gregory:1993vy}. The endpoint of this instability has been argued to be a ten-dimensional localised black hole, which has been constructed in \cite{Dias:2016eto} and also admits a regular Euclidean section. 

This onset of the above instability signals the existence of a zero-mode for the associated Euclidean section. Fixing the energy thus defines a mode that gives a zero-eigenvalue in the matrix of 2nd derivatives of the horizon area.  We expect this eigenvalue to become negative after the onset of the instability.\footnote{\label{foot:instab}Indeed, in Lorentz signature ref. \cite{Hollands:2012sf} shows instabilities to imply that the second variation of bifurcation-surface area under fixed-energy perturbations cannot be negative definite (so that $-\frac{A}{4G}$ cannot have positive definite variations).  And under the expectation that kinetic energy is positive, it is enough to check time-independent perturbations in which we can interpret this $A$ as the area of an extremal suface (and which we can also Wick-rotated to Euclidean signature). }
However, to study this phenomenon in detail, one must resort to analysing fluctuations of the Euclidean Schwarzschild-AdS$\times S^5$ in type IIB supergravity. This is a formidable task that we leave for future endeavours. 

Finally, while the results of our Euclidean analysis seem quite satisfactory, as usual in Euclidean gravity it remains to better understand the proper treatment of the conformal mode.  In particular, while the rule of thumb proposed in section \ref{sec:SL} seems likely to generalize to a wide array of systems, it makes use of both the de Donder gauge and the interpretation of $\hat {\cal G}$ as a metric on the space of perturbations.  Yet there is no clear reason why either of these structures should be physically preferred.  One should thus investigate further the extent to which the results depend on these choices or, even better, to find a first-principles derivation of the correct recipe.  In accord with previous suggestions from \cite{Hartle:2020glw,Schleich:1987fm,Mazur:1989by,Marolf:1996gb,Dasgupta:2001ue,Ambjorn:2002gr,Feldbrugge:2017kzv,Feldbrugge:2017fcc,Feldbrugge:2017mbc}, we expect the Lorentz-signature path integral to provide a useful starting point for such an analysis since, as an oscillatory integral, it should be well-defined in a distribution sense without any Wick rotation.  Allowed deformations of the contour of integration for this path integral might then be used to define the correct Euclidean prescription. We hope to at least partially address this issue in future work.

\paragraph{Acknowledgments}
It is a pleasure to thank Raghu Mahajan for discussions during the initial phase of this work. D.~M. was supported by NSF grants PHY-1801805 and PHY-2107939,  and by funds from the University of California. J.~E.~S has been partially supported by STFC consolidated grants ST/P000681/1, ST/T000694/1

\appendix

\section{\label{app:1}General form for $\delta I^{(2)}_{\partial}$}

We start with the Einstein-Hilbert action with the Gibbons-Hawking-York term for a spacetime $(\mathcal{M},g)$:
\begin{equation}
I = -\frac{1}{16\pi G} \int_\mathcal{M}\mathrm{d}^{d}x\sqrt{g}\left(R-2\Lambda\right)-\frac{1}{8\pi G}\int_{\partial \mathcal{M}}\mathrm{d}^{d-1}x\sqrt{\gamma}\,K \, ,
\end{equation}
with $K$ computed using an outward pointing unit normal $n$, and $\gamma$ the induced metric on $\partial \mathcal{M}$.

Upon variation we find
\begin{equation}
\delta I = \frac{1}{16\pi G} \int_\mathcal{M}\mathrm{d}^{d}x\sqrt{g}\,E^{ab}\,h_{ab}+\frac{1}{16\pi G}\int_{\partial \mathcal{M}}\mathrm{d}^{d-1}x\sqrt{\gamma}\,T^{\mu\nu}\,\alpha_{\mu\nu} \, ,
\end{equation}
where Greek indices run over the boundary $\partial\mathcal{M}$, the object $T^{\mu \nu}$ is the Brown-York stress energy tensor given by
\begin{equation}
T^{\mu\nu}=K^{\mu\nu}-K\gamma^{\mu\nu}\,,
\end{equation}
and we have defined
\begin{equation}
E^{ab}\equiv R^{ab}-\frac{1}{2}R\,g^{ab}+\Lambda g_{ab}\,,\quad h_{ab}\equiv \delta g_{ab}\quad\text{and}\quad \alpha_{\mu\nu}=\delta \gamma_{\mu\nu}\,.
\end{equation}

To find the second order varition we need to take an additional functional derivative to find
\begin{multline}
\delta^2I = \frac{1}{32\pi G} \int_\mathcal{M}\mathrm{d}^{d}x\sqrt{g}\,E^{ab}\,h_{ab}\,h-\frac{1}{8\pi G} \int_\mathcal{M}\mathrm{d}^{d}x\sqrt{g}\, E^{a}_{\phantom{a}c}\,h^{cb}\,h_{ab}
\\
+\frac{1}{16\pi G} \int_\mathcal{M}\mathrm{d}^{d}x\sqrt{g}\, \delta E_{ab}\,h^{ab}
\\
+\frac{1}{32\pi G}\int_{\partial \mathcal{M}}\mathrm{d}^{d-1}x\sqrt{\gamma}\, T^{\mu\nu} \alpha_{\mu\nu}\,\alpha-\frac{1}{8\pi G}\int_{\partial \mathcal{M}}\mathrm{d}^{d-1}x\sqrt{\gamma}\,T^{\mu}_{\phantom{\mu}\rho}\alpha^{\rho\nu}\, \alpha_{\mu\nu}
\\
+\frac{1}{16\pi G}\int_{\partial \mathcal{M}}\mathrm{d}^{d-1}x\sqrt{\gamma}\,\delta T_{\mu\nu}\,\alpha^{\mu\nu}\,.
\label{eq:refc}
\end{multline}
The third term relates to the \L operator
\begin{equation}
\delta E_{ab}=\frac{1}{2}\mathcal{G}_{ab}^{\phantom{ab}cd} \left[(\Delta_L h)_{cd}+2\,\nabla_{(c}\nabla^p\bar{h}_{b)p}\right].
\end{equation}
If we focus on backgrounds that satisfy the equations of motion $E_{ab}=0$, the last three terms in Eq.~(\ref{eq:refc})  provide all the boundary terms to include in our calculation.

\bibliographystyle{jhep}
	\cleardoublepage

\renewcommand*{\bibname}{References}

\bibliography{negative}

\providecommand{\href}[2]{#2}\begingroup\raggedright\begin{thebibliography}{10}

\bibitem{Gibbons:1976ue}
G.~W. Gibbons and S.~W. Hawking, \emph{{Action Integrals and Partition
  Functions in Quantum Gravity}},
  \href{http://dx.doi.org/10.1103/PhysRevD.15.2752}{\emph{Phys. Rev. D}
  {\bfseries 15} (1977) 2752--2756}.

\bibitem{Gibbons:1978ji}
G.~W. Gibbons and M.~J. Perry, \emph{{Quantizing Gravitational Instantons}},
  \href{http://dx.doi.org/10.1016/0550-3213(78)90434-0}{\emph{Nucl. Phys. B}
  {\bfseries 146} (1978) 90--108}.

\bibitem{Allen:1984bp}
B.~Allen, \emph{{Euclidean Schwarzschild negative mode}},
  \href{http://dx.doi.org/10.1103/PhysRevD.30.1153}{\emph{Phys. Rev. D}
  {\bfseries 30} (1984) 1153--1157}.

\bibitem{Prestidge:1999uq}
T.~Prestidge, \emph{{Dynamic and thermodynamic stability and negative modes in
  Schwarzschild-anti-de Sitter}},
  \href{http://dx.doi.org/10.1103/PhysRevD.61.084002}{\emph{Phys. Rev. D}
  {\bfseries 61} (2000) 084002},
  [\href{https://arxiv.org/abs/hep-th/9907163}{{\ttfamily hep-th/9907163}}].

\bibitem{Dasgupta:2001ue}
A.~Dasgupta and R.~Loll, \emph{{A Proper time cure for the conformal sickness
  in quantum gravity}},
  \href{http://dx.doi.org/10.1016/S0550-3213(01)00227-9}{\emph{Nucl. Phys. B}
  {\bfseries 606} (2001) 357--379},
  [\href{https://arxiv.org/abs/hep-th/0103186}{{\ttfamily hep-th/0103186}}].

\bibitem{Kol:2006ga}
B.~Kol, \emph{{The Power of Action: The Derivation of the Black Hole Negative
  Mode}}, \href{http://dx.doi.org/10.1103/PhysRevD.77.044039}{\emph{Phys. Rev.
  D} {\bfseries 77} (2008) 044039},
  [\href{https://arxiv.org/abs/hep-th/0608001}{{\ttfamily hep-th/0608001}}].

\bibitem{Headrick:2006ti}
M.~Headrick and T.~Wiseman, \emph{{Ricci flow and black holes}},
  \href{http://dx.doi.org/10.1088/0264-9381/23/23/006}{\emph{Class. Quant.
  Grav.} {\bfseries 23} (2006) 6683--6708},
  [\href{https://arxiv.org/abs/hep-th/0606086}{{\ttfamily hep-th/0606086}}].

\bibitem{Monteiro:2008wr}
R.~Monteiro and J.~E. Santos, \emph{{Negative modes and the thermodynamics of
  Reissner-Nordstrom black holes}},
  \href{http://dx.doi.org/10.1103/PhysRevD.79.064006}{\emph{Phys. Rev. D}
  {\bfseries 79} (2009) 064006},
  [\href{https://arxiv.org/abs/0812.1767}{{\ttfamily 0812.1767}}].

\bibitem{Monteiro:2009tc}
R.~Monteiro, M.~J. Perry and J.~E. Santos, \emph{{Thermodynamic instability of
  rotating black holes}},
  \href{http://dx.doi.org/10.1103/PhysRevD.80.024041}{\emph{Phys. Rev. D}
  {\bfseries 80} (2009) 024041},
  [\href{https://arxiv.org/abs/0903.3256}{{\ttfamily 0903.3256}}].

\bibitem{Monteiro:2009ke}
R.~Monteiro, M.~J. Perry and J.~E. Santos, \emph{{Semiclassical instabilities
  of Kerr-AdS black holes}},
  \href{http://dx.doi.org/10.1103/PhysRevD.81.024001}{\emph{Phys. Rev. D}
  {\bfseries 81} (2010) 024001},
  [\href{https://arxiv.org/abs/0905.2334}{{\ttfamily 0905.2334}}].

\bibitem{Anninos:2012ft}
D.~Anninos, F.~Denef and D.~Harlow, \emph{{Wave function of
  Vasiliev\textquoteright{}s universe: A few slices thereof}},
  \href{http://dx.doi.org/10.1103/PhysRevD.88.084049}{\emph{Phys. Rev. D}
  {\bfseries 88} (2013) 084049},
  [\href{https://arxiv.org/abs/1207.5517}{{\ttfamily 1207.5517}}].

\bibitem{Benjamin:2020mfz}
N.~Benjamin, S.~Collier and A.~Maloney, \emph{{Pure Gravity and Conical
  Defects}}, \href{http://dx.doi.org/10.1007/JHEP09(2020)034}{\emph{JHEP}
  {\bfseries 09} (2020) 034},
  [\href{https://arxiv.org/abs/2004.14428}{{\ttfamily 2004.14428}}].

\bibitem{Cotler:2019nbi}
J.~Cotler, K.~Jensen and A.~Maloney, \emph{{Low-dimensional de Sitter quantum
  gravity}}, \href{http://dx.doi.org/10.1007/JHEP06(2020)048}{\emph{JHEP}
  {\bfseries 06} (2020) 048},
  [\href{https://arxiv.org/abs/1905.03780}{{\ttfamily 1905.03780}}].

\bibitem{Marolf:2018ldl}
D.~Marolf, \emph{{Microcanonical Path Integrals and the Holography of small
  Black Hole Interiors}},
  \href{http://dx.doi.org/10.1007/JHEP09(2018)114}{\emph{JHEP} {\bfseries 09}
  (2018) 114}, [\href{https://arxiv.org/abs/1808.00394}{{\ttfamily
  1808.00394}}].

\bibitem{Cotler:2021cqa}
J.~Cotler and K.~Jensen, \emph{{Wormholes and black hole microstates in
  AdS/CFT}},  \href{https://arxiv.org/abs/2104.00601}{{\ttfamily 2104.00601}}.

\bibitem{Brown:1992bq}
J.~D. Brown and J.~W. York, Jr., \emph{{The Microcanonical functional integral.
  1. The Gravitational field}},
  \href{http://dx.doi.org/10.1103/PhysRevD.47.1420}{\emph{Phys. Rev. D}
  {\bfseries 47} (1993) 1420--1431},
  [\href{https://arxiv.org/abs/gr-qc/9209014}{{\ttfamily gr-qc/9209014}}].

\bibitem{Brown:1993ke}
J.~D. Brown and J.~W. York, Jr., \emph{{Microcanonical action and the entropy
  of a rotating black hole}},
  \href{http://dx.doi.org/10.1007/978-94-011-1938-2_3}{\emph{Math. Phys. Stud.}
  {\bfseries 15} (1994) 23--34},
  [\href{https://arxiv.org/abs/gr-qc/9303012}{{\ttfamily gr-qc/9303012}}].

\bibitem{Saad:2018bqo}
P.~Saad, S.~H. Shenker and D.~Stanford, \emph{{A semiclassical ramp in SYK and
  in gravity}},  \href{https://arxiv.org/abs/1806.06840}{{\ttfamily
  1806.06840}}.

\bibitem{Regge:1957td}
T.~Regge and J.~A. Wheeler, \emph{{Stability of a Schwarzschild singularity}},
  \href{http://dx.doi.org/10.1103/PhysRev.108.1063}{\emph{Phys. Rev.}
  {\bfseries 108} (1957) 1063--1069}.

\bibitem{Zerilli:1970se}
F.~J. Zerilli, \emph{{Effective potential for even parity Regge-Wheeler
  gravitational perturbation equations}},
  \href{http://dx.doi.org/10.1103/PhysRevLett.24.737}{\emph{Phys. Rev. Lett.}
  {\bfseries 24} (1970) 737--738}.

\bibitem{Hollands:2012sf}
S.~Hollands and R.~M. Wald, \emph{{Stability of Black Holes and Black Branes}},
  \href{http://dx.doi.org/10.1007/s00220-012-1638-1}{\emph{Commun. Math. Phys.}
  {\bfseries 321} (2013) 629--680},
  [\href{https://arxiv.org/abs/1201.0463}{{\ttfamily 1201.0463}}].

\bibitem{CanoPaper}
D.~Marolf and J.~E. Santos, \emph{{The Canonical Ensemble Reloaded: The
  Complex-Stability of Euclidean quantum gravity for Black Holes in a Box}},
  2022.

\bibitem{Witten:2018lgb}
E.~Witten, \emph{{A Note On Boundary Conditions In Euclidean Gravity}},
  \href{https://arxiv.org/abs/1805.11559}{{\ttfamily 1805.11559}}.

\bibitem{Andrade:2015qea}
T.~Andrade, W.~R. Kelly and D.~Marolf, \emph{{Einstein\textendash{}Maxwell
  Dirichlet walls, negative kinetic energies, and the adiabatic approximation
  for extreme black holes}},
  \href{http://dx.doi.org/10.1088/0264-9381/32/19/195017}{\emph{Class. Quant.
  Grav.} {\bfseries 32} (2015) 195017},
  [\href{https://arxiv.org/abs/1503.03915}{{\ttfamily 1503.03915}}].

\bibitem{Andrade:2015gja}
T.~Andrade, W.~R. Kelly, D.~Marolf and J.~E. Santos, \emph{{On the stability of
  gravity with Dirichlet walls}},
  \href{http://dx.doi.org/10.1088/0264-9381/32/23/235006}{\emph{Class. Quant.
  Grav.} {\bfseries 32} (2015) 235006},
  [\href{https://arxiv.org/abs/1504.07580}{{\ttfamily 1504.07580}}].

\bibitem{Dong:2019piw}
X.~Dong and D.~Marolf, \emph{{One-loop universality of holographic codes}},
  \href{http://dx.doi.org/10.1007/JHEP03(2020)191}{\emph{JHEP} {\bfseries 03}
  (2020) 191}, [\href{https://arxiv.org/abs/1910.06329}{{\ttfamily
  1910.06329}}].

\bibitem{Banados:1993qp}
M.~Banados, C.~Teitelboim and J.~Zanelli, \emph{{Black hole entropy and the
  dimensional continuation of the Gauss-Bonnet theorem}},
  \href{http://dx.doi.org/10.1103/PhysRevLett.72.957}{\emph{Phys. Rev. Lett.}
  {\bfseries 72} (1994) 957--960},
  [\href{https://arxiv.org/abs/gr-qc/9309026}{{\ttfamily gr-qc/9309026}}].

\bibitem{Gibbons:1978ac}
G.~W. Gibbons, S.~W. Hawking and M.~J. Perry, \emph{{Path Integrals and the
  Indefiniteness of the Gravitational Action}},
  \href{http://dx.doi.org/10.1016/0550-3213(78)90161-X}{\emph{Nucl. Phys. B}
  {\bfseries 138} (1978) 141--150}.

\bibitem{Gratton:1999ya}
S.~Gratton and N.~Turok, \emph{{Cosmological perturbations from the no boundary
  Euclidean path integral}},
  \href{http://dx.doi.org/10.1103/PhysRevD.60.123507}{\emph{Phys. Rev. D}
  {\bfseries 60} (1999) 123507},
  [\href{https://arxiv.org/abs/astro-ph/9902265}{{\ttfamily
  astro-ph/9902265}}].

\bibitem{Gratton:2000fj}
S.~Gratton and N.~Turok, \emph{{Homogeneous modes of cosmological instantons}},
  \href{http://dx.doi.org/10.1103/PhysRevD.63.123514}{\emph{Phys. Rev. D}
  {\bfseries 63} (2001) 123514},
  [\href{https://arxiv.org/abs/hep-th/0008235}{{\ttfamily hep-th/0008235}}].

\bibitem{Gratton:2001gw}
S.~Gratton, A.~Lewis and N.~Turok, \emph{{Closed universes from cosmological
  instantons}}, \href{http://dx.doi.org/10.1103/PhysRevD.65.043513}{\emph{Phys.
  Rev. D} {\bfseries 65} (2002) 043513},
  [\href{https://arxiv.org/abs/astro-ph/0111012}{{\ttfamily
  astro-ph/0111012}}].

\bibitem{Gross:1982cv}
D.~J. Gross, M.~J. Perry and L.~G. Yaffe, \emph{{Instability of Flat Space at
  Finite Temperature}},
  \href{http://dx.doi.org/10.1103/PhysRevD.25.330}{\emph{Phys. Rev. D}
  {\bfseries 25} (1982) 330--355}.

\bibitem{Dias:2009iu}
O.~J.~C. Dias, P.~Figueras, R.~Monteiro, J.~E. Santos and R.~Emparan,
  \emph{{Instability and new phases of higher-dimensional rotating black
  holes}}, \href{http://dx.doi.org/10.1103/PhysRevD.80.111701}{\emph{Phys. Rev.
  D} {\bfseries 80} (2009) 111701},
  [\href{https://arxiv.org/abs/0907.2248}{{\ttfamily 0907.2248}}].

\bibitem{Dias:2010eu}
O.~J.~C. Dias, P.~Figueras, R.~Monteiro, H.~S. Reall and J.~E. Santos,
  \emph{{An instability of higher-dimensional rotating black holes}},
  \href{http://dx.doi.org/10.1007/JHEP05(2010)076}{\emph{JHEP} {\bfseries 05}
  (2010) 076}, [\href{https://arxiv.org/abs/1001.4527}{{\ttfamily 1001.4527}}].

\bibitem{Kudoh:2006bp}
H.~Kudoh, \emph{{Origin of black string instability}},
  \href{http://dx.doi.org/10.1103/PhysRevD.73.104034}{\emph{Phys. Rev. D}
  {\bfseries 73} (2006) 104034},
  [\href{https://arxiv.org/abs/hep-th/0602001}{{\ttfamily hep-th/0602001}}].

\bibitem{Dias:2015nua}
O.~J.~C. Dias, J.~E. Santos and B.~Way, \emph{{Numerical Methods for Finding
  Stationary Gravitational Solutions}},
  \href{http://dx.doi.org/10.1088/0264-9381/33/13/133001}{\emph{Class. Quant.
  Grav.} {\bfseries 33} (2016) 133001},
  [\href{https://arxiv.org/abs/1510.02804}{{\ttfamily 1510.02804}}].

\bibitem{Freund:1980xh}
P.~G.~O. Freund and M.~A. Rubin, \emph{{Dynamics of Dimensional Reduction}},
  \href{http://dx.doi.org/10.1016/0370-2693(80)90590-0}{\emph{Phys. Lett. B}
  {\bfseries 97} (1980) 233--235}.

\bibitem{Martinec:strings98}
E.~Martinec, ``The d-star and its decays.'' 1998.

\bibitem{Banks:1998dd}
T.~Banks, M.~R. Douglas, G.~T. Horowitz and E.~J. Martinec, \emph{{AdS dynamics
  from conformal field theory}},
  \href{https://arxiv.org/abs/hep-th/9808016}{{\ttfamily hep-th/9808016}}.

\bibitem{Peet:1998cr}
A.~W. Peet and S.~F. Ross, \emph{{Microcanonical phases of string theory on
  AdS(m) x S**n}},
  \href{http://dx.doi.org/10.1088/1126-6708/1998/12/020}{\emph{JHEP} {\bfseries
  12} (1998) 020}, [\href{https://arxiv.org/abs/hep-th/9810200}{{\ttfamily
  hep-th/9810200}}].

\bibitem{Hubeny:2002xn}
V.~E. Hubeny and M.~Rangamani, \emph{{Unstable horizons}},
  \href{http://dx.doi.org/10.1088/1126-6708/2002/05/027}{\emph{JHEP} {\bfseries
  05} (2002) 027}, [\href{https://arxiv.org/abs/hep-th/0202189}{{\ttfamily
  hep-th/0202189}}].

\bibitem{Dias:2015pda}
O.~J.~C. Dias, J.~E. Santos and B.~Way, \emph{{Lumpy AdS$_{5}$\texttimes{}
  S$^{5}$ black holes and black belts}},
  \href{http://dx.doi.org/10.1007/JHEP04(2015)060}{\emph{JHEP} {\bfseries 04}
  (2015) 060}, [\href{https://arxiv.org/abs/1501.06574}{{\ttfamily
  1501.06574}}].

\bibitem{Buchel:2015gxa}
A.~Buchel and L.~Lehner, \emph{{Small black holes in $AdS_5\times S^5$}},
  \href{http://dx.doi.org/10.1088/0264-9381/32/14/145003}{\emph{Class. Quant.
  Grav.} {\bfseries 32} (2015) 145003},
  [\href{https://arxiv.org/abs/1502.01574}{{\ttfamily 1502.01574}}].

\bibitem{Dias:2016eto}
O.~J.~C. Dias, J.~E. Santos and B.~Way, \emph{{Localised $AdS_5\times S^5$
  Black Holes}},
  \href{http://dx.doi.org/10.1103/PhysRevLett.117.151101}{\emph{Phys. Rev.
  Lett.} {\bfseries 117} (2016) 151101},
  [\href{https://arxiv.org/abs/1605.04911}{{\ttfamily 1605.04911}}].

\bibitem{Cardona:2020unx}
B.~Cardona and P.~Figueras, \emph{{Critical lumpy black holes in AdS${}_p\times
  S^q$}}, \href{http://dx.doi.org/10.1007/JHEP05(2021)265}{\emph{JHEP}
  {\bfseries 21} (2020) 265},
  [\href{https://arxiv.org/abs/2103.06932}{{\ttfamily 2103.06932}}].

\bibitem{Gregory:1993vy}
R.~Gregory and R.~Laflamme, \emph{{Black strings and p-branes are unstable}},
  \href{http://dx.doi.org/10.1103/PhysRevLett.70.2837}{\emph{Phys. Rev. Lett.}
  {\bfseries 70} (1993) 2837--2840},
  [\href{https://arxiv.org/abs/hep-th/9301052}{{\ttfamily hep-th/9301052}}].

\bibitem{Hartle:2020glw}
J.~B. Hartle and K.~Schleich, \emph{{The Conformal Rotation in Linearised
  Gravity}},  in \emph{Quantum Field Theory and Quantum Statistics} (C.~J.~I.
  I.~A.~Batalin and G.~A. Vilkovisky, eds.), pp.~{67--87}, 4, 1987.
\newblock \href{https://arxiv.org/abs/2004.06635}{{\ttfamily 2004.06635}}.

\bibitem{Schleich:1987fm}
K.~Schleich, \emph{{Conformal Rotation in Perturbative Gravity}},
  \href{http://dx.doi.org/10.1103/PhysRevD.36.2342}{\emph{Phys. Rev. D}
  {\bfseries 36} (1987) 2342--2363}.

\bibitem{Mazur:1989by}
P.~O. Mazur and E.~Mottola, \emph{{The Gravitational Measure, Solution of the
  Conformal Factor Problem and Stability of the Ground State of Quantum
  Gravity}}, \href{http://dx.doi.org/10.1016/0550-3213(90)90268-I}{\emph{Nucl.
  Phys. B} {\bfseries 341} (1990) 187--212}.

\bibitem{Marolf:1996gb}
D.~Marolf, \emph{{Path integrals and instantons in quantum gravity:
  Minisuperspace models}},
  \href{http://dx.doi.org/10.1103/PhysRevD.53.6979}{\emph{Phys. Rev. D}
  {\bfseries 53} (1996) 6979--6990},
  [\href{https://arxiv.org/abs/gr-qc/9602019}{{\ttfamily gr-qc/9602019}}].

\bibitem{Ambjorn:2002gr}
J.~Ambjorn, A.~Dasgupta, J.~Jurkiewicz and R.~Loll, \emph{{A Lorentzian cure
  for Euclidean troubles}},
  \href{http://dx.doi.org/10.1016/S0920-5632(01)01903-X}{\emph{Nucl. Phys. B
  Proc. Suppl.} {\bfseries 106} (2002) 977--979},
  [\href{https://arxiv.org/abs/hep-th/0201104}{{\ttfamily hep-th/0201104}}].

\bibitem{Feldbrugge:2017kzv}
J.~Feldbrugge, J.-L. Lehners and N.~Turok, \emph{{Lorentzian Quantum
  Cosmology}}, \href{http://dx.doi.org/10.1103/PhysRevD.95.103508}{\emph{Phys.
  Rev. D} {\bfseries 95} (2017) 103508},
  [\href{https://arxiv.org/abs/1703.02076}{{\ttfamily 1703.02076}}].

\bibitem{Feldbrugge:2017fcc}
J.~Feldbrugge, J.-L. Lehners and N.~Turok, \emph{{No smooth beginning for
  spacetime}},
  \href{http://dx.doi.org/10.1103/PhysRevLett.119.171301}{\emph{Phys. Rev.
  Lett.} {\bfseries 119} (2017) 171301},
  [\href{https://arxiv.org/abs/1705.00192}{{\ttfamily 1705.00192}}].

\bibitem{Feldbrugge:2017mbc}
J.~Feldbrugge, J.-L. Lehners and N.~Turok, \emph{{No rescue for the no boundary
  proposal: Pointers to the future of quantum cosmology}},
  \href{http://dx.doi.org/10.1103/PhysRevD.97.023509}{\emph{Phys. Rev. D}
  {\bfseries 97} (2018) 023509},
  [\href{https://arxiv.org/abs/1708.05104}{{\ttfamily 1708.05104}}].

\end{thebibliography}\endgroup

\end{document}